\documentclass[twoside,11pt]{article} 
\usepackage{jmlr2e}

\jmlrheading{17}{2016}{1-17}{11/2015; Revised 10/2016}{10/2016}{Micha\"el Chichignoud, Johannes Lederer, Martin J. Wainwright}
\ShortHeadings{Lasso Tuning}{Chichignoud, Lederer, and Wainwright}
\firstpageno{1}

\usepackage{pdfsync}

\usepackage{graphicx}
\usepackage{amssymb}
\usepackage{algorithm2e}

\newcommand{\GoodEvent}[1]{\ensuremath{\mathcal{T}_{#1}}}

\newcommand{\Xmat}{\ensuremath{X}}
\newcommand{\pdim}{\ensuremath{p}}
\newcommand{\numobs}{\ensuremath{n}}
\newcommand{\defn}{\ensuremath{: \, = }}
\newcommand{\betalasso}{{\widehat\beta_\lambda}}

\newcommand{\KEYCONTWO}{\ensuremath{\widebar{C}}}
\newcommand{\KEYCON}{\ensuremath{C}}

\newcommand{\lamhat}{\ensuremath{\hat{\lambda}}}

\newcommand{\betahat}{\ensuremath{\widehat{\beta}}}
\newcommand{\delpar}{\ensuremath{\delta}}

\newcommand{\uj}{\sqrt\frac{\log(p)}{n}}
\newcommand{\ujs}{\sqrt{\log(p)/n}}

\newlength{\widebarargwidth}
\newlength{\widebarargheight}
\newlength{\widebarargdepth}
\DeclareRobustCommand{\widebar}[1]{%
  \settowidth{\widebarargwidth}{\ensuremath{#1}}%
  \settoheight{\widebarargheight}{\ensuremath{#1}}%
  \settodepth{\widebarargdepth}{\ensuremath{#1}}%
  \addtolength{\widebarargwidth}{-0.3\widebarargheight}%
  \addtolength{\widebarargwidth}{-0.3\widebarargdepth}%
  \makebox[0pt][l]{\hspace{0.3\widebarargheight}%
    \hspace{0.3\widebarargdepth}%
    \addtolength{\widebarargheight}{0.3ex}%
    \rule[\widebarargheight]{0.95\widebarargwidth}{0.1ex}}%
  {#1}}

\newcommand{\GoodEventStar}{\mathcal{T}^*_{\delta}}

\usepackage{color}

\newcommand{\betastar}{\ensuremath{\beta^*}}
\newcommand{\real}{\ensuremath{\mathbb{R}}}

\newcommand{\that}{\ensuremath{\widehat{t}}}

\newcommand{\lammax}{\ensuremath{\lambda_{\operatorname{max}}}}

\newcommand{\kdim}{\ensuremath{\tilde s}}

\newcommand{\Ssettrue}{\ensuremath{S}}
\newcommand{\Sbartrue}{\ensuremath{\Ssettrue^c}}

\newcommand{\Cone}{\ensuremath{\mathbb{C}}}

\newcommand{\spindex}{\kdim}
\newcommand{\SamCov}{\ensuremath{\widehat{\Sigma}}}

\newcommand{\HACKCON}{\ensuremath{4}}
\newcommand{\DelHat}{\ensuremath{\widehat{\Delta}}}
\newcommand{\zhat}{\ensuremath{\widehat{z}}}

\newcommand{\PWINC}[1]{\ensuremath{\rho(#1)}}

\newcommand{\inprod}[2]{\ensuremath{\langle #1 , \, #2 \rangle}}

\input{Notation.sty}

\begin{document}

\newcommand{\tnew}[1]{\textcolor{red}{#1}}
\newcommand{\told}[1]{\textcolor{olive}{#1}}
\newcommand{\tcomment}[1]{\textcolor{blue}{#1}}

\newcommand{\jaj}[1]{#1}

\title{\jaj{A Practical Scheme and Fast Algorithm to\\Tune the Lasso With
  Optimality Guarantees}}

\author{\name Micha\"el Chichignoud \email
  michael.chichignoud@gmail.com\\ \addr Seminar for Statistics\\ ETH
  Z\"urich \AND \name Johannes Lederer\thanks{Corresponding
    author. Postal address: Department of Statistics at University of
    Washington, Box~354322, Seattle, WA 98195} \email
  ledererj@uw.edu\\ \addr Department of Statistics\\ and Department of
  Biostatistics\\ University of Washington \AND Martin J. Wainwright
  \email wainwrig@berkeley.edu\\ \addr Department of Statistics\\ and
  Department of Electrical Engineering and Computer
  Sciences\\ University of California at Berkeley }

\editor{Francis Bach}
\maketitle




\begin{abstract}%
We introduce a novel scheme for choosing the regularization parameter
in high-dimensional linear regression with Lasso. This scheme,
inspired by Lepski's method for bandwidth selection in non-parametric
regression, is equipped with both optimal finite-sample guarantees and
a fast algorithm. In particular, for any design matrix such that the
Lasso has low sup-norm error under an ``oracle choice'' of the
regularization parameter, we show that our method matches the oracle
performance up to a small constant factor, and show that it can be
implemented by performing simple tests along a single Lasso path.  By
applying the Lasso to simulated and real data, \jaj{we find that our
  novel scheme can be faster and more accurate than standard schemes
  such as Cross-Validation.}
\end{abstract}

\newcommand{\revision}[1]{#1}


\section{Introduction}

Regularized estimators---among them the
Lasso~\citep{Tibshirani-LASSO}, the Square-Root and the Scaled Lasso
\citep{Antoniadis10,Belloni11,Stadler10,ScaledLasso11}, as well as
estimators based on nonconvex penalties such as MCP~\citep{Zhang10}
and SCAD~\citep{Fan_Li01}---all hinge on finding a ``suitable'' choice
of tuning parameters.  There are many possible methods for solving
this so-called calibration problem, but for high-dimensional
regression problems, there is a not a single method that is
computationally tractable and for which the non-asymptotic theory is
well understood.

The focus of this paper is the calibration of the Lasso for sparse
linear regression, where the tuning parameter needs to be adjusted to
both the noise distribution and the design
matrix~\citep{vdGeer11,YoyoMomo12,ArnakMoYo14}. Calibration schemes
for this setting are typically based on Cross-Validation (CV) or
BIC-type criteria. However, CV-based procedures can be computationally
intensive and are currently lacking in non-asymptotic theory for
high-dimensional problems.  BIC-type criteria, on the other hand, are
computationally simpler but also lacking in non-asymptotic
guarantees. Another approach is to replace the Lasso with Square-Root
Lasso or TREX~\citep{Lederer:14}; however, Square-Root Lasso still
contains a tuning parameter that needs to be calibrated to certain
aspects of the model, and the theory for TREX is currently
fragmentary.  For these reasons and given the extensive use of the
Lasso in practice, understanding the calibration of Lasso is
important.

In this paper, we introduce a new scheme for calibrating the Lasso in
the supremum norm ($\ell_\infty$)-loss, which we refer to as
\emph{Adaptive Validation for $\ell_\infty$} (\avi).  This method is
based on tests that are inspired by Lepski's method for non-parametric
regression~\citep{Lepski90,Lepski_Mammen_Spokoiny97}, see
also~\cite{ChichiYoyo12}. In contrast to current schemes for the
Lasso, our method is equipped with both optimal theoretical guarantees
and a fast computational routine.

The remainder of this paper is organized as follows. In
Section~\ref{methodology}, we introduce the \avi~method.  Our main
theoretical results show that this method enjoys finite sample
guarantees for the calibration of Lasso with respect to sup-norm loss
(Theorem~\ref{ThmAvi}) and \jaj{variable selection
  (Remark~\ref{RemElli}). }  In addition, we provide a simple and fast
algorithm (Algorithm~\ref{foxalgo}).  In Section~\ref{simulations}, we
illustrate these features with applications to simulated data and to
biological data.  We conclude with a discussion in
Section~\ref{conclusions}.

\emph{Notation:} The indicator of events is denoted by
$\1\{\cdot\}\in\{0,1\}$, the cardinality of sets by $|\cdot|$, the
sup-norm or maximum norm of vectors in $\R^p$ vectors
$\|\cdot\|_\infty$, the number of non-zero entries by $\|\cdot\|_0$,
the $\ell_1$- and $\ell_2$-norms by $\|\cdot\|_1$ and $\|\cdot\|_2$,
respectively, and $[p]\defn\{1,\dots,p\}$. For given vector $\beta \in
\rp$ and subset $A$ of $[p]$, $\beta_{A}\in\R^{|A|}$ and
$\beta_{A^c}\in\R^{|A^c|}$ denote the components in $A$ and in its
complement~$A^c$, respectively.

\section{Background and Methodology}
\label{methodology}

In this section, we introduce some background and then move onto a
description of the \avi~method.


\subsection{Framework}

We study the calibration of the Lasso tuning parameter in
high-dimensional linear regression models that can contain many
predictors and allow for the possibility of correlated and
heavy-tailed noise. More specifically, we assume that the data $(Y,X)$
with outcome $Y\in\rn$ and design matrix $X\in\rnp$ is distributed
according to a linear regression model
\begin{equation}
\label{model}\tag{Model}
Y = X \bt+\varepsilon,
\end{equation}
where $\bt \in \rp$ is the regression vector and $\varepsilon\in\rn$
is a random noise vector.  Our framework allows for $p$ larger than
$n$ and requires that the noise variables $\varepsilon$ satisfy only
the second moment condition
\begin{align}
\label{EqnSecondMoment}
\max \limits_{i\in\otn} \E[\varepsilon_{i}^2]<\infty.
\end{align}
A standard approach for estimating $\bt$ in such a model is by
computing the $\ell_1$-regularized least-squares estimate, known as
the Lasso, and given by
\begin{equation}\tag{Lasso}
 \label{Lasso}
\widehat\beta_\lambda\in\argmin_{\beta\in\R^p}\left\{\frac{\|Y-X\beta\|_2^2}{n}
+ \lambda \|\beta\|_1\right\}.
\end{equation}
Note that this equation actually defines a family of estimators
indexed by the tuning parameter $\lambda>0$, which determines the
level of regularization.

Intuitively, the optimal choice of $\lambda$ is dictated by a
trade-off between bias and some form of variance control.  Bias is
induced by the shrinkage effect of the $\ell_1$-regularizer, which
acts even on non-zero coordinates of the regression vector.  Thus, the
bias grows as $\lambda$ is increased.  On the other hand,
$\ell_1$-regularization is useful in canceling out fluctuations in the
score function, which for the linear regression model is given by
${\Xmat^\top \varepsilon}/{\numobs}$.  Thus, an optimal choice of
$\lambda$ is the smallest one that is large enough to control these
fluctuations.

A large body of theoretical work (e.g.,~\cite{Sara09,
  Bickel09,Buhlmann11,NegRavWaiYu12}) has shown that an appropriate
formalization of this intuition is based on the event
\begin{align}
\label{EqnDefnGoodEvent}
\GoodEvent{\lambda} & \defn \Big \{ \frac{\|\Xmat^\top
  \varepsilon\|_\infty}{\numobs} \leq \frac{\lambda}{\HACKCON} \Big\}.
\end{align}
When this event holds, then as long as the design matrix $\Xmat$ is
``well-behaved'', it is possible to obtain bounds on the sup-norm
error of the Lasso estimate.  There are various ways of characterizing
well-behaved design matrices; of most relevance for sup-norm error
control are mutual incoherence conditions~\citep{BuneaEN,Lounici08} as
well as $\ell_\infty$-restricted eigenvalues~\citep{YeZhang10}.
See~\cite{Sara09} and Section~\ref{SecCompatibility} for further
discussion of these design conditions.

In order to bring sharp focus to the calibration problem, rather than
focusing on any particular design condition, it is useful to instead
work under the generic assumption that the Lasso sup-norm error is
controlled under the event $\GoodEvent{\lambda}$ defined in
equation~\eqref{EqnDefnGoodEvent}.  More formally, we state:

\begin{assumption}[$\ell_\infty(\KEYCON)$] There is a numerical constant 
$\KEYCON$ such that conditioned on $\GoodEvent{\lambda}$, the Lasso
  $\ell_\infty$-error is upper bounded as $\|\betalasso -
  \betastar\|_\infty \leq \KEYCON \lambda$.
\end{assumption}
\noindent As mentioned above, there are many conditions on the design
matrix $\Xmat$ under which Assumption~$\ell_\infty(\KEYCON)$ is valid,
and we consider a number of them in the sequel.

With this set-up in place, we can now focus specifically on how to
choose the regularization parameter.  Since we can handle only
finitely many tuning parameters in practice, we restrict ourselves to
the selection of a tuning parameter among a finite but arbitrarily
large number of choices.  It is easy to see that $\lammax \defn 2
\|X^\top Y\|_\infty/\numobs$ is the smallest tuning parameter for
which $\betahat_\lambda$ equals zero.  Accordingly, for a given
positive integer $N\in\N$, let us form the grid 
\begin{align*}
0 < \lambda_1 < \cdots < \lambda_N = \lammax,
\end{align*}
denoted by $\Lambda \defn \{\lambda_1,\dots, \lambda_N\}$ for
short. Assumption $\ell_\infty(\KEYCON)$ guarantees that the sup-norm
error is proportional to $\lambda$ whenever the event
$\GoodEvent{\lambda}$ holds; consequently, for a given probability of
error $\delpar \in (0,1)$, it is natural to choose the smallest
$\lambda$ for which event $\GoodEvent{\lambda}$ holds with probability
at least $1 - \delpar$, assuming that it is finite. This criterion can
be formalized as follows:
\begin{definition}[Oracle tuning parameter]
\label{DefnOracleParameter}
For any constant $\delpar \in (0,1)$, the oracle tuning parameter is
given by
\begin{align}
\label{EqnOracleParameter}
    \ot\defn \arg \min_{\lambda\in\Lambda}\left\{\mpr\left(\mathcal
    T_\lambda\right)\geq 1-\delpar\right\}.
\end{align}
\end{definition}
\noindent Note that by construction, if we solve the Lasso using the
oracle choice $\ot$, and if the design matrix~$X$ fulfills Assumption
$\ell_\infty(\KEYCON)$, then the resulting estimate satisfies the
bound \mbox{$\|\betahat_{\ot} - \betastar\|_\infty \leq \KEYCON \ot$}
with probability at least $1-\delta$.  Unfortunately, the oracle
choice is inaccessible to us, since we cannot compute the probability
of the event $\GoodEvent{\lambda}$ based on the observed data.
However, as we now describe, we can mimic    
this performance, up to a factor of three, using a simple data-dependent procedure.


\subsection{Adaptive Calibration Scheme}

Let us now describe a data-dependent scheme for choosing the
regularization parameter, referred to as Adaptive Calibration for
$\ell_\infty$ (\avi):
\begin{definition}[\avi]\label{dtp} Under Assumption $\ell_\infty(\KEYCON)$ and for a given constant $\KEYCONTWO\geq \KEYCON$, Adaptive Calibration for $\ell_\infty$ (\avi) selects the tuning parameter
\newcommand{\suchthat}{\;\ifnum\currentgrouptype=16 \middle\fi|\;}
\begin{align}\label{AVLambdaDefn}
  \lamhat & \defn \min \Bigg\{ \lambda\in\Lambda \,\Big\rvert\,
  \max_{\substack{\lambda',\lambda''\in\Lambda\\\lambda',\lambda''\geq \lambda}} \Bigg[\frac{
      \|\betahat_{\lambda'}- \betahat_{\lambda''}\|_\infty}{
      \lambda'+\lambda''}- \KEYCONTWO \Bigg] \leq 0\Bigg\}.
\end{align}
\end{definition}
\noindent The definition is based on tests for sup-norm
differences of Lasso estimates with different tuning parameters. We stress that Definition~\ref{dtp}
requires neither prior knowledge about the regression vector nor about the
noise.

The tests in Definition~\ref{dtp} can be formulated in terms of the
binary random variables
\begin{align*}
\that_{\lambda_j} & \defn \prod_{k=j}^N
\1\left\{\frac{\|\betahat_{\lambda_j} -
  \betahat_{\lambda_k}\|_\infty}{\lambda_j+\lambda_k}- \KEYCONTWO \leq
0\right\} \qquad \mbox{for $j \in [N]$,}
\end{align*}
from the \avi~tuning parameter $\lamhat$ can be computed as follows:

\begin{algorithm}[H]\label{foxalgo}
  \KwData{$\betahat_{\lambda_1},\dots,\betahat_{\lambda_N},\KEYCONTWO$}
  \KwResult{$\lamhat\in \Lambda$}
\vspace{0.3cm}Set initial index: $j\leftarrow N$\\
\vspace{0.15cm}\While{$\that_{\lambda_{j-1}}\neq 0$ and $j>1$}{
  Update index: $j\leftarrow j-1$\\ } Set output:
$\lamhat\leftarrow\lambda_{j}$
\vspace{0.3cm}
 \caption{Algorithm for \avi~in Definition~\ref{dtp}.}
\end{algorithm}
\noindent This algorithm can be readily implemented and only requires
the computation of one Lasso solution path.  In strong contrast,
$k$-fold Cross-Validation requires the computation of $k$ solution
paths. Consequently, the Lasso with \avi~can be computed about $k$
times faster than Lasso with $k$-fold Cross-Validation.

The following result guarantees that the Lasso with \avi~method
achieves the sup-norm error up to a constant pre-factor:
\begin{theorem}[Optimality of \avi~]
\label{ThmAvi} 
Suppose that condition $\ell_\infty(\KEYCON)$ holds and the
\avi~method is implemented with parameter $\KEYCONTWO \geq \KEYCON$.
Then for any $\delpar \in (0,1)$, the \avi~output pair $(\lamhat,
\betahat_{\lamhat})$ given by the rule~\eqref{AVLambdaDefn} satisfies
the bounds
\begin{align}
\label{EqnAVIMain}
\lamhat \leq \ot \quad \mbox{and} \quad
\|\betahat_{\lamhat}-\bt\|_\infty \, \leq 3 \KEYCONTWO \ot
\end{align}
with probability at least $1-\delpar$.
\end{theorem}
\jaj{\begin{remark}[Relevance for estimation and variable
      selection]\label{RemElli} The $\ell_\infty$-bound from
    equation~\eqref{EqnAVIMain} directly implies that the \avi\ scheme
    is adaptively optimal for the estimation of the regression
    vector~$\bt$ in $\ell_\infty$-loss. As another important feature,
    Theorem~\ref{ThmAvi} entails strong {variable selection
      guarantees}. First, the $\ell_\infty$-bound implies that
    \avi\ recovers all non-zero entries of the regression vector $\bt$
    that are larger than $3\KEYCONTWO\ot$ in absolute
    value. Additionally, by virtue of the bound~$\lamhat \leq \ot,$
    thresholding $\betahat_{\lamhat}$ by $3\KEYCONTWO\lamhat$ leads to
    exact support recovery if all non-zero entries of $\bt$ are larger
    than $6\KEYCONTWO\ot$ in absolute value. In strong contrast,
    standard calibration schemes are not equipped with comparable
    variable selection guarantees, and there is no theoretically sound
    guidance for how to threshold standard schemes.
\end{remark}}

We prove Theorem~\ref{ThmAvi} in Appendix~\ref{AppThmAvi}; here let us
make a few remarks about its consequences.  First, if we knew the
oracle value $\ot$ defined in equation~\eqref{EqnOracleParameter},
then under Assumption~$\ell_\infty(\KEYCON)$, the Lasso estimate
$\betahat$ would satisfy the $\ell_\infty$-bound $\|\betahat -
\bt\|_\infty \leq \KEYCON \ot$.  Consequently, when the \avi~method is
implemented with parameter $\KEYCON$, then its sup-norm error is
optimal up to a factor of three. For standard calibration schemes,
among them Cross-Validation, no comparable guarantees are available in
the literature. In fact, we are not aware of \emph{any} finite sample
guarantees for standard calibration schemes.

We point out that Theorem~\ref{ThmAvi}---in contrast to asymptotic
results or results with unspecified constants---provides explicit
guarantees for arbitrary sample sizes. Moreover, Theorem~\ref{ThmAvi}
does not presume prior knowledge about the regression vector or the
noise distribution and allows, in particular, for correlated,
heavy-tailed noise.  From the perspective of theoretical sharpness,
the best choice for $\KEYCONTWO$ is $\KEYCONTWO=\KEYCON$. However,
Theorem~\ref{ThmAvi} shows that it also suffices to know an upper
bound for $\KEYCON$.  We provide more details on choices of $\KEYCON$
and $\KEYCONTWO$ below.

We finally observe that the specific choice of the grid enters
Theorem~\ref{ThmAvi} only via the oracle. Indeed, for any choice of
the grid, Theorem~\ref{ThmAvi} ensures that $\lamhat$ performs as well
as the oracle tuning parameter~$\ot$, which is the ``best'' tuning
parameter on the grid.

\subsection{Conditions on the Design Matrix for $\ell_\infty$-guarantees}
\label{SecCompatibility}

Let us now describe some  conditions on the design matrix
$\Xmat$ that are sufficient for Assumption~$\ell_\infty(\KEYCON)$.  \jaj{We stress that these are conditions to ensure that the \emph{Lasso} satisfies $\ell_\infty$-bounds; importantly, our method itself does not impose any additional restrictions. } We defer all proofs of the results stated here to
Appendix~\ref{AppSecCompatibility} and, for simplicity, we assume in the following that the sample covariance $\SamCov \defn
{\Xmat^\top \Xmat}/{\numobs}$ has been normalized
such that $\SamCov_{jj} = 1$ for all $j \in [\pdim]$.

The significance of the event $\GoodEvent{\lambda}$ lies in the
following implication: when $\GoodEvent{\lambda}$ holds, then it can
be shown (e.g.,\cite{Bickel09,Buhlmann11,NegRavWaiYu12}) that the
Lasso error $\DelHat \defn \betalasso - \betastar$ must belong to the
cone
\begin{align}\label{ConeDefn}
\Cone(\Ssettrue) & \defn \big \{ \Delta \in \real^\pdim \, \mid \,
\|\Delta_{\Sbartrue} \|_1 \leq 2 \|\Delta_{\Ssettrue}\|_1 \big \},
\end{align}
where $\Ssettrue$ denotes the support of $\betastar$, and $\Sbartrue$
its complement.  Accordingly, all known conditions involve controlling
the behavior of the sample covariance matrix~$\SamCov$ for vectors
lying within this cone.

The most directly stated sufficient condition is based on lower
bounding the \emph{$\ell_\infty$-restricted eigenvalue:} there exists
some $\gamma > 0$ such that
\begin{align}
\label{EqnGammaRE}
\|\SamCov \Delta\|_\infty & \geq \gamma \|\Delta\|_\infty \qquad
\mbox{for all $\Delta \in \Cone(\Ssettrue)$.}
\end{align}
See~\cite{Sara09} for an overview of various conditions for the Lasso,
and their relations. Based on~\eqref{EqnGammaRE}, we prove in
Appendix~\ref{ProofLemma1} the following result:

\begin{lemma}[$\ell_\infty$-restricted eigenvalue]
\label{LemGammaRE}
Suppose that $\SamCov$ satisfies the $\gamma$-RE
condition~\eqref{EqnGammaRE} and that $\GoodEvent{\lambda}$ holds.
 Then Assumption~$\ell_\infty(\KEYCON)$ is valid  with $\KEYCON = \frac{5}{4 \gamma}$.
\end{lemma}
Although this result is cleanly stated, the RE condition cannot be
verified in practice, since it involves the unknown support set
$\Ssettrue$.  Accordingly, let us now state some sufficient and verifiable
conditions for obtaining bounds on the restricted eigenvalues, and
hence for verifying Assumption~$\ell_\infty(\KEYCON)$.

For a given integer $\spindex \in [2, \pdim]$ and scalar $\nu > 0$,
let us say that the sample covariance $\SamCov$ is diagonally dominant
with parameters $(\spindex, \nu)$ if
\begin{align}
\label{EqnDiagonalDominant}
\max_{ \substack{|T| = \spindex \\ T \subset [\pdim] \backslash
    \{j\}}} \sum_{k \in T } |\SamCov_{jk}| & < \nu \qquad \mbox{for
  all $j \in [\pdim]$.}
\end{align}
In the context of this definition, the reader should recall that we
have assumed that $\SamCov_{jj} = 1$ for all $j \in [\pdim]$.  Note
that this condition can be verified in polynomial-time, since the
subset $T$ achieving the maximum in row $j$ can be obtained simply by
sorting the entries $\{|\SamCov_{jk}|, k \in [\pdim] \backslash j \}$.
The significance of this condition lies in the following result:
\begin{lemma}[Diagonal dominance of order $\spindex$]
\label{LemDiagonalDominant}
Suppose that $\spindex\geq 9|\Ssettrue|$ and $\SamCov$ is $\spindex$-order diagonally dominant with
parameter $\nu \in [0,1)$.  Then under the event $\GoodEvent{\lambda}$,
  Assumption $\ell_\infty(\KEYCON)$ is valid with $\KEYCON =
  \frac{5}{4 (1 -\nu)}$.
\end{lemma}
\noindent See Appendix~\ref{ProofLemma2} for the proof.

It is worth noting that the diagonal dominance condition is weaker
than the pairwise incoherence conditions that have been used in past
work on sup-norm error~\citep{Lounici08}.  The pairwise incoherence of
the sample covariance is given by $\PWINC{\SamCov} = \max_{j \neq k}
|\SamCov_{jk}|$. If the pairwise incoherence satisfies the bound
$\PWINC{\SamCov} \leq {\nu}/{\spindex}$, then it follows that
$\SamCov$ is diagonally dominant with parameters~$(\spindex, \nu)$.

 By combining Lemma~\ref{LemDiagonalDominant} with
Theorem~\ref{ThmAvi}, we obtain the following corollary:
\begin{corollary}
\label{CorDiagonallyDominant}
Suppose that $\spindex\geq 9 |\Ssettrue|$ and $\SamCov$ is
$\spindex$-order diagonally dominant with parameter $\nu \in [0, 1)$.
  Then for any $\delpar \in (0,1)$, the \avi~ method with $\KEYCONTWO
  = \frac{5}{4 \, (1 - \nu)}$ returns an estimate $\betahat_{\lamhat}$
  such that
\begin{align}
\|\betahat_{\lamhat} - \betastar\|_\infty \leq \frac{15}{4 \, (1 -
  \nu)} \ot
\end{align}
with probability at least $1-\delpar$.
\end{corollary}

Another sufficient condition for the sup-norm optimality of \avi is a
design compatibility condition due to~\cite{vandeG07}.  For each index
$j\in \otp$, suppose that we define the deterministic vector
\begin{equation*}
\eta^j \in \arg \min_{\substack{\beta\in\R^{p} \\\beta_j=-1}} \left
\{\frac{\|X\beta\|_2^2}{n} + \uj \|\beta\|_1 \right \}.
\end{equation*}
Note that this optimization problem defining the vector regression of
the $j$th column of the design matrix on the set of all other columns,
where we have imposed an $\ell_1$-penalty with weight $\uj$.  We can
then derive the following sup-norm bound for the Lasso.

\begin{lemma}[Lasso bound under compatibility]\label{multlem}
 Assume that $X$ fulfills the compatibility condition
\begin{equation}\tag{Compatibility}
\min_{\|\beta_{\Sbartrue}\|_1\leq
  3\|\beta_{\Ssettrue}\|_1}\left\{\frac{\sqrt{|\Ssettrue|}\|X\beta\|_2}{\sqrt
  n\|\beta_{\Ssettrue}\|_1}\right\} \; \geq t
\end{equation}
for a constant $t>0$. Additionally, assume that
\begin{align*}
\sup_{j\in\otp} \frac{|\Ssettrue|}{{t^{2}}\|\eta^j\|_1}\leq
\frac{1}{\log n}\sqrt\frac{n}{\log p}.
\end{align*}
Then under the event $\GoodEvent{\lambda}$, Assumption
$\ell_\infty(\KEYCON)$ is valid with
\begin{equation*}
\KEYCON \defn \left(\frac{3}{4}+\frac{1}{\log
  (n)}\right)\max_{j\in[p]}\frac{\|\eta^j\|_1}{\|X\eta^j\|_2^2/n+\ujs\|\eta^j\|_{-j}/2}.
  \end{equation*}
\end{lemma}
\noindent This bound is a consequence of results in \citep{vdGeer14};
the proof is deferred to Section~\ref{sec:proofsmultfox}.  We are now
ready to state the optimality of \avi~with respect to this bound.
\begin{corollary}[Optimality of \avi]
\label{CorAvim}
Assume that the assumptions in Lemma~\ref{multlem} are met. Then for
any constant $\delta>0$, the following bound for Lasso \avi~with
$\KEYCONTWO=\KEYCON$, and $\KEYCON$ as above, holds with probability
at least $1-\delta$:
\begin{equation}
\|\betahat_{\lamhat}-\bt\|_\infty\leq 3C \ot.
  \end{equation}
\end{corollary}
\noindent This result demonstrates the optimality of \avi~for sup-norm loss under the compatibility condition.
\begin{remark}[Constant $\KEYCONTWO$ \revision{in practice}]\label{motivationC} 
\revision{The optimal choice is $\KEYCONTWO=\KEYCON$ in view of our
  theoretical results. The constant~$\KEYCON$ (or an upper bound of
  it) can be readily computed, because it depends only on~$X$
  (cf. Lemma~\ref{multlem}) or on $X$ and an upper bound on $s$
  (cf. Lemma~\ref{LemDiagonalDominant}). However, we propose the
  universal choice $\KEYCON=0.75$ for all practical purposes. Note
  that accurate support recovery and $\ell_\infty$-estimation is
  possible only if the design is near orthogonal. A direct computation
  yields the bound $\|\betalasso - \betastar\|_\infty \leq \KEYCON
  \lambda$ with $\KEYCON=0.75$ for orthogonal design. Letting
  $\alpha\to\infty$ in Theorem 1 due to~\cite{Lounici08} yields the
  same bound with $\KEYCON\approx 0.75$ for near orthogonal
  designs. This provides strong theoretical support for the choice
  $\KEYCONTWO=0.75$. The empirical evidence in
  Section~\ref{simulations} indicates that a further calibration is
  indeed not necessary.}
\end{remark}

\section{Simulations}
\label{simulations}

\newcommand{\resultbestO}{0.19} \newcommand{\resultbestE}{0.04}
\newcommand{\fpresultbestO}{0} \newcommand{\fpresultbestE}{0}
\newcommand{\fnresultbestO}{0} \newcommand{\fnresultbestE}{0}
\newcommand{\resultaviO}{0.22}
\newcommand{\resultaviE}{0.03}
\newcommand{\fpresultaviO}{0.51}
\newcommand{\fpresultaviE}{0.93}
\newcommand{\fnresultaviO}{0.00}
\newcommand{\fnresultaviE}{0.00}
\newcommand{\resultcvO}{0.23}
\newcommand{\resultcvE}{0.05}
\newcommand{\fpresultcvO}{26.96}
\newcommand{\fpresultcvE}{15.90}
\newcommand{\fnresultcvO}{0}
\newcommand{\fnresultcvE}{0}

\newcommand{\resultbestOO}{0.21}
\newcommand{\resultbestEE}{0.04}
\newcommand{\fpresultbestOO}{0}
\newcommand{\fpresultbestEE}{0}
\newcommand{\fnresultbestOO}{0}
\newcommand{\fnresultbestEE}{0}
\newcommand{\resultaviOO}{0.23}
\newcommand{\resultaviEE}{0.04}
\newcommand{\fpresultaviOO}{0.3}
\newcommand{\fpresultaviEE}{0.85}
\newcommand{\fnresultaviOO}{0.06}
\newcommand{\fnresultaviEE}{0.51}
\newcommand{\resultcvOO}{0.25}
\newcommand{\resultcvEE}{0.06}
\newcommand{\fpresultcvOO}{27.27}
\newcommand{\fpresultcvEE}{15.43}
\newcommand{\fnresultcvOO}{0}
\newcommand{\fnresultcvEE}{0}

\newcommand{\resultbestOOO}{0.24}
\newcommand{\resultbestEEE}{0.05}
\newcommand{\fpresultbestOOO}{0}
\newcommand{\fpresultbestEEE}{0}
\newcommand{\fnresultbestOOO}{0}
\newcommand{\fnresultbestEEE}{0}
\newcommand{\resultaviOOO}{0.30}
\newcommand{\resultaviEEE}{0.08}
\newcommand{\fpresultaviOOO}{0.02}
\newcommand{\fpresultaviEEE}{0.14}
\newcommand{\fnresultaviOOO}{2.57}
\newcommand{\fnresultaviEEE}{2.72}
\newcommand{\resultcvOOO}{0.29}
\newcommand{\resultcvEEE}{0.07}
\newcommand{\fpresultcvOOO}{29.42}
\newcommand{\fpresultcvEEE}{15.9}
\newcommand{\fnresultcvOOO}{0}
\newcommand{\fnresultcvEEE}{0}

\newcommand{\SresultbestO}{0.22}
\newcommand{\SresultbestE}{0.05}
\newcommand{\SfpresultbestO}{0}
\newcommand{\SfpresultbestE}{0}
\newcommand{\SfnresultbestO}{0}
\newcommand{\SfnresultbestE}{0}
\newcommand{\SresultaviO}{0.25}
\newcommand{\SresultaviE}{0.05}
\newcommand{\SfpresultaviO}{15.83}
\newcommand{\SfpresultaviE}{14.84}
\newcommand{\SfnresultaviO}{0}
\newcommand{\SfnresultaviE}{0}
\newcommand{\SresultcvO}{0.26}
\newcommand{\SresultcvE}{0.05}
\newcommand{\SfpresultcvO}{42.26}
\newcommand{\SfpresultcvE}{24.14}
\newcommand{\SfnresultcvO}{0}
\newcommand{\SfnresultcvE}{0}

\newcommand{\SresultbestOO}{0.25}
\newcommand{\SresultbestEE}{0.06}
\newcommand{\SfpresultbestOO}{0}
\newcommand{\SfpresultbestEE}{0}
\newcommand{\SfnresultbestOO}{0}
\newcommand{\SfnresultbestEE}{0}
\newcommand{\SresultaviOO}{0.28}
\newcommand{\SresultaviEE}{0.07}
\newcommand{\SfpresultaviOO}{4.51}
\newcommand{\SfpresultaviEE}{5.88}
\newcommand{\SfnresultaviOO}{0.0}
\newcommand{\SfnresultaviEE}{0.0}
\newcommand{\SresultcvOO}{0.29}
\newcommand{\SresultcvEE}{0.07}
\newcommand{\SfpresultcvOO}{40.48}
\newcommand{\SfpresultcvEE}{26.03}
\newcommand{\SfnresultcvOO}{0}
\newcommand{\SfnresultcvEE}{0}

\newcommand{\SresultbestOOO}{0.29}
\newcommand{\SresultbestEEE}{0.07}
\newcommand{\SfpresultbestOOO}{0}
\newcommand{\SfpresultbestEEE}{0}
\newcommand{\SfnresultbestOOO}{0}
\newcommand{\SfnresultbestEEE}{0}
\newcommand{\SresultaviOOO}{0.33}
\newcommand{\SresultaviEEE}{0.08}
\newcommand{\SfpresultaviOOO}{0.2}
\newcommand{\SfpresultaviEEE}{1.26}
\newcommand{\SfnresultaviOOO}{2.42}
\newcommand{\SfnresultaviEEE}{2.64}
\newcommand{\SresultcvOOO}{0.34}
\newcommand{\SresultcvEEE}{0.07}
\newcommand{\SfpresultcvOOO}{39.38}
\newcommand{\SfpresultcvEEE}{23.57}
\newcommand{\SfnresultcvOOO}{0}
\newcommand{\SfnresultcvEEE}{0}

\newcommand{\correlations}{\kappa}
\newcommand{\coloravi}{red!50!white}
\newcommand{\colorcv}{blue!40!white}
\newcommand{\colorbest}{green!40!white}

\newcommand{\factor}{35}
\newcommand{\shift}{-2.5}
\newcommand{\doubleshift}{2*\shift}

\newcommand{\vsfactor}{0.20}
\newcommand{\vsdoubleshift}{-5}
\newcommand{\vsshift}{-1.25}
\newcommand{\vszeroshift}{2.5}

\newcommand{\nameplotoracle}{Oracle}

\newcommand{\pone}{\ensuremath{300}}
\newcommand{\ptwo}{\ensuremath{900}}
\newcommand{\signaltonoise}{\ensuremath{5}}
\newcommand{\numberlambda}{\ensuremath{{100}}}
\newcommand{\lambdaexponent}{\ensuremath{1.3}}
\newcommand{\sparsity}{\ensuremath{6}}
\newcommand{\kappaone}{\ensuremath{0}}
\newcommand{\kappatwo}{\ensuremath{0.2}}
\newcommand{\kappathree}{\ensuremath{0.4}}

\newcommand{\captionsup}{Sup-norm error $\|\widehat\beta_\lambda-\bt\|_\infty$ of the Lasso with three different calibration schemes for the tuning parameter $\lambda$. Depicted are the results for three simulation settings that differ in the correlation level $\kappa.$ The simulation settings and the calibration schemes are specified in the body of the text.}

\newcommand{\captionvs}{Number of false positives $|\{j:\bt_j= 0,(\widehat\beta_\lambda)_j\neq 0\}|$ and false negatives $|\{j:\bt_j\neq 0,(\widehat\beta_\lambda)_j= 0\}|$ of the Lasso with \avi\ and Cross-Validation as calibration schemes for the tuning parameter~$\lambda$. For \avi, the safe threshold described after Theorem~\ref{ThmAvi} is applied. The simulations settings correspond to those in Figure~\ref{MainComparison1}.}

\newcommand{\captionvst}{Number of false positives $|\{j:\bt_j= 0,(\widehat\beta_\lambda)_j\neq 0\}|$ and false negatives $|\{j:\bt_j\neq 0,(\widehat\beta_\lambda)_j= 0\}|$ of the Lasso with \avi\ and Cross-Validation as calibration schemes for the tuning parameter~$\lambda$. For \avi, the safe threshold described after Theorem~\ref{ThmAvi} is applied. The simulations settings correspond to those in Figure~\ref{MainComparison2}.}

In this section, we perform experiments on both simulated and
real data to demonstrate the practical performance of \avi.


\subsection{Simulated Data}

We simulate data from linear regression models as in
equation~\eqref{model} with $n=200$ observations and
$p\in\{\pone,\ptwo\}$ parameters. More specifically, we sample each
row of the design matrix $X\in\R^{n\times p}$ from a $p$-dimensional
normal distribution with mean $0$ and covariance matrix
$(1-\kappa)\operatorname{I}+\kappa\1$, where $\operatorname{I}$ is the
identity matrix, $\1\defn(1,\dots,1)^\top(1,\dots,1)$ is the matrix of
ones, and $\correlations\in\{\kappaone,\kappatwo,\kappathree\}$ is the
magnitude of the mutual correlations. For the entries of the noise
$\varepsilon\in\rn$, we take the one-dimensional normal distribution
with mean $0$ and variance~$1$. The entries of $\bt$ are first set to
$0$ except for~\sparsity\ uniformly at random chosen entries that are
each set to $1$ or $-1$ with equal probability. The entire vector
$\bt$ is then rescaled such that the signal-to-noise ratio ${\|X
  \bt\|_2^2}/{n}$ is equal to~\signaltonoise. We finally consider a
grid of~\numberlambda\ tuning parameters
$\Lambda:=\{\lambda_{\max}/\lambdaexponent^0,
\lambda_{\max}/\lambdaexponent^1,\dots,
\lambda_{\max}/\lambdaexponent^{99}\}$ with
$\lambda_{\max}\defn2\|X^\top Y\|_\infty/n$.  We run $100$~experiments
for each set of parameters and report the corresponding means (thick,
colored bars) and standard deviations (thin, black lines). All
computations are conducted with the software~R~\citep{Rsoftware} and
the glmnet package~\citep{glmnet10}. \jaj{While we restrict the
  presentation to the parameter settings described, we found similar
  results over a wide range of settings.}


\begin{figure}
\begin{tikzpicture}

\path [\colorbest,fill=\colorbest, line width = 0pt] (0,1+\doubleshift) rectangle (\resultbestOOO*\factor,1.5+\doubleshift);
\draw (\resultbestOOO*\factor-\resultbestEEE*\factor,1.25+\doubleshift) -| (\resultbestOOO*\factor+\resultbestEEE*\factor,1.25+\doubleshift);
\draw (\resultbestOOO*\factor-\resultbestEEE*\factor, 1.4+\doubleshift) -- (\resultbestOOO*\factor-\resultbestEEE*\factor, 1.1+\doubleshift);
\draw (\resultbestOOO*\factor+\resultbestEEE*\factor, 1.4+\doubleshift) -- (\resultbestOOO*\factor+\resultbestEEE*\factor, 1.1+\doubleshift);
\filldraw (\resultbestOOO*\factor,1.25+\doubleshift) circle (2pt);

\path [\coloravi,fill=\coloravi, line width = 0pt] (0,0.5+\doubleshift) rectangle (\resultaviOOO*\factor,1+\doubleshift);
\draw (\resultaviOOO*\factor-\resultaviEEE*\factor,0.75+\doubleshift) -| (\resultaviOOO*\factor+\resultaviEEE*\factor,0.75+\doubleshift);
\draw (\resultaviOOO*\factor-\resultaviEEE*\factor, 0.9+\doubleshift) -- (\resultaviOOO*\factor-\resultaviEEE*\factor, 0.6+\doubleshift);
\draw (\resultaviOOO*\factor+\resultaviEEE*\factor, 0.9+\doubleshift) -- (\resultaviOOO*\factor+\resultaviEEE*\factor, 0.6+\doubleshift);
\filldraw (\resultaviOOO*\factor,0.75+\doubleshift) circle (2pt);

\path [\colorcv,fill=\colorcv, line width = 0pt] (0,0+\doubleshift) rectangle (\resultcvOOO*\factor,.5+\doubleshift);
\draw (\resultcvOOO*\factor-\resultcvEEE*\factor,0.25+\doubleshift) -| (\resultcvOOO*\factor+\resultcvEEE*\factor,0.25+\doubleshift);
\draw (\resultcvOOO*\factor-\resultcvEEE*\factor, 0.4+\doubleshift) -- (\resultcvOOO*\factor-\resultcvEEE*\factor, 0.1+\doubleshift);
\draw (\resultcvOOO*\factor+\resultcvEEE*\factor, 0.4+\doubleshift) -- (\resultcvOOO*\factor+\resultcvEEE*\factor, 0.1+\doubleshift);
\filldraw (\resultcvOOO*\factor,0.25+\doubleshift) circle (2pt);

\node[text width = 15mm] at (-1,.7+\doubleshift) {$p=\pone$ $\correlations=0.4$};

\node at (2,1.25+\doubleshift) {\nameplotoracle};
\node at (2,0.75+\doubleshift) {\avi};
\node at (2,0.25+\doubleshift) {Cross-Validation};

\path [\colorbest,fill=\colorbest, line width = 0pt] (0,1+\shift) rectangle (\resultbestOO*\factor,1.5+\shift);
\draw (\resultbestOO*\factor-\resultbestEE*\factor,1.25+\shift) -| (\resultbestOO*\factor+\resultbestEE*\factor,1.25+\shift);
\draw (\resultbestOO*\factor-\resultbestEE*\factor, 1.4+\shift) -- (\resultbestOO*\factor-\resultbestEE*\factor, 1.1+\shift);
\draw (\resultbestOO*\factor+\resultbestEE*\factor, 1.4+\shift) -- (\resultbestOO*\factor+\resultbestEE*\factor, 1.1+\shift);
\filldraw (\resultbestOO*\factor,1.25+\shift) circle (2pt);

\path [\coloravi,fill=\coloravi, line width = 0pt] (0,0.5+\shift) rectangle (\resultaviOO*\factor,1+\shift);
\draw (\resultaviOO*\factor-\resultaviEE*\factor,0.75+\shift) -| (\resultaviOO*\factor+\resultaviEE*\factor,0.75+\shift);
\draw (\resultaviOO*\factor-\resultaviEE*\factor, 0.9+\shift) -- (\resultaviOO*\factor-\resultaviEE*\factor, 0.6+\shift);
\draw (\resultaviOO*\factor+\resultaviEE*\factor, 0.9+\shift) -- (\resultaviOO*\factor+\resultaviEE*\factor, 0.6+\shift);
\filldraw (\resultaviOO*\factor,0.75+\shift) circle (2pt);

\path [\colorcv,fill=\colorcv, line width = 0pt] (0,0+\shift) rectangle (\resultcvOO*\factor,.5+\shift);
\draw (\resultcvOO*\factor-\resultcvEE*\factor,0.25+\shift) -| (\resultcvOO*\factor+\resultcvEE*\factor,0.25+\shift);
\draw (\resultcvOO*\factor-\resultcvEE*\factor, 0.4+\shift) -- (\resultcvOO*\factor-\resultcvEE*\factor, 0.1+\shift);
\draw (\resultcvOO*\factor+\resultcvEE*\factor, 0.4+\shift) -- (\resultcvOO*\factor+\resultcvEE*\factor, 0.1+\shift);
\filldraw (\resultcvOO*\factor,0.25+\shift) circle (2pt);

\node[text width = 15mm] at (-1,.7+\shift) {$p=\pone$ $\correlations=0.2$};

\node at (2,1.25+\shift) {\nameplotoracle};
\node at (2,0.75+\shift) {\avi};
\node at (2,0.25+\shift) {Cross-Validation};

\path [\colorbest,fill=\colorbest, line width = 0pt] (0,1) rectangle (\resultbestO*\factor,1.5);
\draw (\resultbestO*\factor-\resultbestE*\factor,1.25) -| (\resultbestO*\factor+\resultbestE*\factor,1.25);
\draw (\resultbestO*\factor-\resultbestE*\factor, 1.4) -- (\resultbestO*\factor-\resultbestE*\factor, 1.1);
\draw (\resultbestO*\factor+\resultbestE*\factor, 1.4) -- (\resultbestO*\factor+\resultbestE*\factor, 1.1);
\filldraw (\resultbestO*\factor,1.25) circle (2pt);

\path [\coloravi,fill=\coloravi, line width = 0pt] (0,0.5) rectangle (\resultaviO*\factor,1);
\draw (\resultaviO*\factor-\resultaviE*\factor,0.75) -| (\resultaviO*\factor+\resultaviE*\factor,0.75);
\draw (\resultaviO*\factor-\resultaviE*\factor, 0.9) -- (\resultaviO*\factor-\resultaviE*\factor, 0.6);
\draw (\resultaviO*\factor+\resultaviE*\factor, 0.9) -- (\resultaviO*\factor+\resultaviE*\factor, 0.6);
\filldraw (\resultaviO*\factor,0.75) circle (2pt);

\path [\colorcv,fill=\colorcv, line width = 0pt] (0,0) rectangle (\resultcvO*\factor,.5);
\draw (\resultcvO*\factor-\resultcvE*\factor,0.25) -| (\resultcvO*\factor+\resultcvE*\factor,0.25);
\draw (\resultcvO*\factor-\resultcvE*\factor, 0.4) -- (\resultcvO*\factor-\resultcvE*\factor, 0.1);
\draw (\resultcvO*\factor+\resultcvE*\factor, 0.4) -- (\resultcvO*\factor+\resultcvE*\factor, 0.1);
\filldraw (\resultcvO*\factor,0.25) circle (2pt);

\node[text width = 15mm] at (-1,.7) {$p=\pone$ $\correlations=0$};

\node at (2,1.25) {\nameplotoracle};
\node at (2,0.75) {\avi};
\node at (2,0.25) {Cross-Validation};

\draw[line width = 1pt, black!80, dotted] (0.1*\factor,-5.8) --(0.1*\factor,2);
\draw[line width = 1pt, black!80, dotted] (0.2*\factor,-5.8) --(0.2*\factor,2);
\draw[line width = 1pt, black!80, dotted] (0.3*\factor,-5.8) --(0.3*\factor,2);

\draw[line width = 3pt] (-0,-5.8) --(0,2); 
\draw[line width = 3pt, ->] (-0.3,-5.5) --(14,-5.5); 
\node at (14,-6.2) {$\ell_\infty$ error}; 
\node at (0,-6.2) {$0$};
\node at (0.1*\factor,-6.2) {$0.1$};
\node at (0.2*\factor,-6.2) {$0.2$};
\node at (0.3*\factor,-6.2) {$0.3$};

\end{tikzpicture}
\caption{\captionsup}
\label{MainComparison1}
\end{figure}


\begin{figure}
\begin{tikzpicture}

\path [\colorbest,fill=\colorbest, line width = 0pt] (0,1+\doubleshift) rectangle (\SresultbestOOO*\factor,1.5+\doubleshift);
\draw (\SresultbestOOO*\factor-\SresultbestEEE*\factor,1.25+\doubleshift) -| (\SresultbestOOO*\factor+\SresultbestEEE*\factor,1.25+\doubleshift);
\draw (\SresultbestOOO*\factor-\SresultbestEEE*\factor, 1.4+\doubleshift) -- (\SresultbestOOO*\factor-\SresultbestEEE*\factor, 1.1+\doubleshift);
\draw (\SresultbestOOO*\factor+\SresultbestEEE*\factor, 1.4+\doubleshift) -- (\SresultbestOOO*\factor+\SresultbestEEE*\factor, 1.1+\doubleshift);
\filldraw (\SresultbestOOO*\factor,1.25+\doubleshift) circle (2pt);

\path [\coloravi,fill=\coloravi, line width = 0pt] (0,0.5+\doubleshift) rectangle (\SresultaviOOO*\factor,1+\doubleshift);
\draw (\SresultaviOOO*\factor-\SresultaviEEE*\factor,0.75+\doubleshift) -| (\SresultaviOOO*\factor+\SresultaviEEE*\factor,0.75+\doubleshift);
\draw (\SresultaviOOO*\factor-\SresultaviEEE*\factor, 0.9+\doubleshift) -- (\SresultaviOOO*\factor-\SresultaviEEE*\factor, 0.6+\doubleshift);
\draw (\SresultaviOOO*\factor+\SresultaviEEE*\factor, 0.9+\doubleshift) -- (\SresultaviOOO*\factor+\SresultaviEEE*\factor, 0.6+\doubleshift);
\filldraw (\SresultaviOOO*\factor,0.75+\doubleshift) circle (2pt);

\path [\colorcv,fill=\colorcv, line width = 0pt] (0,0+\doubleshift) rectangle (\SresultcvOOO*\factor,.5+\doubleshift);
\draw (\SresultcvOOO*\factor-\SresultcvEEE*\factor,0.25+\doubleshift) -| (\SresultcvOOO*\factor+\SresultcvEEE*\factor,0.25+\doubleshift);
\draw (\SresultcvOOO*\factor-\SresultcvEEE*\factor, 0.4+\doubleshift) -- (\SresultcvOOO*\factor-\SresultcvEEE*\factor, 0.1+\doubleshift);
\draw (\SresultcvOOO*\factor+\SresultcvEEE*\factor, 0.4+\doubleshift) -- (\SresultcvOOO*\factor+\SresultcvEEE*\factor, 0.1+\doubleshift);
\filldraw (\SresultcvOOO*\factor,0.25+\doubleshift) circle (2pt);

\node[text width = 15mm] at (-1,.7+\doubleshift) {$p=\ptwo$ $\correlations=0.4$};

\node at (2,1.25+\doubleshift) {\nameplotoracle};
\node at (2,0.75+\doubleshift) {\avi};
\node at (2,0.25+\doubleshift) {Cross-Validation};

\path [\colorbest,fill=\colorbest, line width = 0pt] (0,1+\shift) rectangle (\SresultbestOO*\factor,1.5+\shift);
\draw (\SresultbestOO*\factor-\SresultbestEE*\factor,1.25+\shift) -| (\SresultbestOO*\factor+\SresultbestEE*\factor,1.25+\shift);
\draw (\SresultbestOO*\factor-\SresultbestEE*\factor, 1.4+\shift) -- (\SresultbestOO*\factor-\SresultbestEE*\factor, 1.1+\shift);
\draw (\SresultbestOO*\factor+\SresultbestEE*\factor, 1.4+\shift) -- (\SresultbestOO*\factor+\SresultbestEE*\factor, 1.1+\shift);
\filldraw (\SresultbestOO*\factor,1.25+\shift) circle (2pt);

\path [\coloravi,fill=\coloravi, line width = 0pt] (0,0.5+\shift) rectangle (\SresultaviOO*\factor,1+\shift);
\draw (\SresultaviOO*\factor-\SresultaviEE*\factor,0.75+\shift) -| (\SresultaviOO*\factor+\SresultaviEE*\factor,0.75+\shift);
\draw (\SresultaviOO*\factor-\SresultaviEE*\factor, 0.9+\shift) -- (\SresultaviOO*\factor-\SresultaviEE*\factor, 0.6+\shift);
\draw (\SresultaviOO*\factor+\SresultaviEE*\factor, 0.9+\shift) -- (\SresultaviOO*\factor+\SresultaviEE*\factor, 0.6+\shift);
\filldraw (\SresultaviOO*\factor,0.75+\shift) circle (2pt);

\path [\colorcv,fill=\colorcv, line width = 0pt] (0,0+\shift) rectangle (\SresultcvOO*\factor,.5+\shift);
\draw (\SresultcvOO*\factor-\SresultcvEE*\factor,0.25+\shift) -| (\SresultcvOO*\factor+\SresultcvEE*\factor,0.25+\shift);
\draw (\SresultcvOO*\factor-\SresultcvEE*\factor, 0.4+\shift) -- (\SresultcvOO*\factor-\SresultcvEE*\factor, 0.1+\shift);
\draw (\SresultcvOO*\factor+\SresultcvEE*\factor, 0.4+\shift) -- (\SresultcvOO*\factor+\SresultcvEE*\factor, 0.1+\shift);
\filldraw (\SresultcvOO*\factor,0.25+\shift) circle (2pt);

\node[text width = 15mm] at (-1,.7+\shift) {$p=\ptwo$ $\correlations=0.2$};

\node at (2,1.25+\shift) {\nameplotoracle};
\node at (2,0.75+\shift) {\avi};
\node at (2,0.25+\shift) {Cross-Validation};

\path [\colorbest,fill=\colorbest, line width = 0pt] (0,1) rectangle (\SresultbestO*\factor,1.5);
\draw (\SresultbestO*\factor-\SresultbestE*\factor,1.25) -| (\SresultbestO*\factor+\SresultbestE*\factor,1.25);
\draw (\SresultbestO*\factor-\SresultbestE*\factor, 1.4) -- (\SresultbestO*\factor-\SresultbestE*\factor, 1.1);
\draw (\SresultbestO*\factor+\SresultbestE*\factor, 1.4) -- (\SresultbestO*\factor+\SresultbestE*\factor, 1.1);
\filldraw (\SresultbestO*\factor,1.25) circle (2pt);

\path [\coloravi,fill=\coloravi, line width = 0pt] (0,0.5) rectangle (\SresultaviO*\factor,1);
\draw (\SresultaviO*\factor-\SresultaviE*\factor,0.75) -| (\SresultaviO*\factor+\SresultaviE*\factor,0.75);
\draw (\SresultaviO*\factor-\SresultaviE*\factor, 0.9) -- (\SresultaviO*\factor-\SresultaviE*\factor, 0.6);
\draw (\SresultaviO*\factor+\SresultaviE*\factor, 0.9) -- (\SresultaviO*\factor+\SresultaviE*\factor, 0.6);
\filldraw (\SresultaviO*\factor,0.75) circle (2pt);

\path [\colorcv,fill=\colorcv, line width = 0pt] (0,0) rectangle (\SresultcvO*\factor,.5);
\draw (\SresultcvO*\factor-\SresultcvE*\factor,0.25) -| (\SresultcvO*\factor+\SresultcvE*\factor,0.25);
\draw (\SresultcvO*\factor-\SresultcvE*\factor, 0.4) -- (\SresultcvO*\factor-\SresultcvE*\factor, 0.1);
\draw (\SresultcvO*\factor+\SresultcvE*\factor, 0.4) -- (\SresultcvO*\factor+\SresultcvE*\factor, 0.1);
\filldraw (\SresultcvO*\factor,0.25) circle (2pt);

\node[text width = 15mm] at (-1,.7) {$p=\ptwo$ $\correlations=0$};

\node at (2,1.25) {\nameplotoracle};
\node at (2,0.75) {\avi};
\node at (2,0.25) {Cross-Validation};

\draw[line width = 1pt, black!80, dotted] (0.1*\factor,-5.8) --(0.1*\factor,2);
\draw[line width = 1pt, black!80, dotted] (0.2*\factor,-5.8) --(0.2*\factor,2);
\draw[line width = 1pt, black!80, dotted] (0.3*\factor,-5.8) --(0.3*\factor,2);

\draw[line width = 3pt] (-0,-5.8) --(0,2); 
\draw[line width = 3pt, ->] (-0.3,-5.5) --(14,-5.5); 
\node at (14,-6.2) {$\ell_\infty$ error}; 
\node at (0,-6.2) {$0$};
\node at (0.1*\factor,-6.2) {$0.1$};
\node at (0.2*\factor,-6.2) {$0.2$};
\node at (0.3*\factor,-6.2) {$0.3$};

\end{tikzpicture}
\caption{\captionsup}
\label{MainComparison2}
\end{figure}


\begin{figure}
\begin{center}
\begin{tikzpicture}
  
\path [\coloravi,fill=\coloravi, line width = 0pt] (0,1.9+\vsdoubleshift) rectangle (\fpresultaviOOO*\vsfactor,2.4+\vsdoubleshift);
\draw (\fpresultaviOOO*\vsfactor-\fpresultaviEEE*\vsfactor,2.15+\vsdoubleshift) -| (\fpresultaviOOO*\vsfactor+\fpresultaviEEE*\vsfactor,2.15+\vsdoubleshift);
\draw (\fpresultaviOOO*\vsfactor-\fpresultaviEEE*\vsfactor, 2.3+\vsdoubleshift) -- (\fpresultaviOOO*\vsfactor-\fpresultaviEEE*\vsfactor, 2+\vsdoubleshift);
\draw (\fpresultaviOOO*\vsfactor+\fpresultaviEEE*\vsfactor, 2.3+\vsdoubleshift) -- (\fpresultaviOOO*\vsfactor+\fpresultaviEEE*\vsfactor, 2+\vsdoubleshift);
\filldraw (\fpresultaviOOO*\vsfactor,2.15+\vsdoubleshift) circle (2pt);
\path [\colorcv,fill=\colorcv, line width = 0pt] (0,1.4+\vsdoubleshift) rectangle (\fpresultcvOOO*\vsfactor,1.9+\vsdoubleshift);
\draw (\fpresultcvOOO*\vsfactor-\fpresultcvEEE*\vsfactor,1.65+\vsdoubleshift) -| (\fpresultcvOOO*\vsfactor+\fpresultcvEEE*\vsfactor,1.65+\vsdoubleshift);
\draw (\fpresultcvOOO*\vsfactor-\fpresultcvEEE*\vsfactor, 1.8+\vsdoubleshift) -- (\fpresultcvOOO*\vsfactor-\fpresultcvEEE*\vsfactor, 1.5+\vsdoubleshift);
\draw (\fpresultcvOOO*\vsfactor+\fpresultcvEEE*\vsfactor, 1.8+\vsdoubleshift) -- (\fpresultcvOOO*\vsfactor+\fpresultcvEEE*\vsfactor, 1.5+\vsdoubleshift);
\filldraw (\fpresultcvOOO*\vsfactor,1.65+\vsdoubleshift) circle (2pt);

\path [\coloravi,fill=\coloravi, line width = 0pt] (0,0.5+\vsdoubleshift) rectangle (\fnresultaviOOO*\vsfactor,1+\vsdoubleshift);
\draw (\fnresultaviOOO*\vsfactor-\fnresultaviEEE*\vsfactor,0.75+\vsdoubleshift) -| (\fnresultaviOOO*\vsfactor+\fnresultaviEEE*\vsfactor,0.75+\vsdoubleshift);
\draw (\fnresultaviOOO*\vsfactor-\fnresultaviEEE*\vsfactor, 0.9+\vsdoubleshift) -- (\fnresultaviOOO*\vsfactor-\fnresultaviEEE*\vsfactor, 0.6+\vsdoubleshift);
\draw (\fnresultaviOOO*\vsfactor+\fnresultaviEEE*\vsfactor, 0.9+\vsdoubleshift) -- (\fnresultaviOOO*\vsfactor+\fnresultaviEEE*\vsfactor, 0.60+\vsdoubleshift);
\filldraw (\fnresultaviOOO*\vsfactor,0.75+\vsdoubleshift) circle (2pt);
\path [\colorcv,fill=\colorcv, line width = 0pt] (0,-0.0+\vsdoubleshift) rectangle (\fnresultcvOOO*\vsfactor,0.5+\vsdoubleshift);
\draw (\fnresultcvOOO*\vsfactor-\fnresultcvEEE*\vsfactor,0.25+\vsdoubleshift) -| (\fnresultcvOOO*\vsfactor+\fnresultcvEEE*\vsfactor,0.25+\vsdoubleshift);
\draw (\fnresultcvOOO*\vsfactor-\fnresultcvEEE*\vsfactor, 0.4+\vsdoubleshift) -- (\fnresultcvOOO*\vsfactor-\fnresultcvEEE*\vsfactor, 0.1+\vsdoubleshift);
\draw (\fnresultcvOOO*\vsfactor+\fnresultcvEEE*\vsfactor, 0.4+\vsdoubleshift) -- (\fnresultcvOOO*\vsfactor+\fnresultcvEEE*\vsfactor, 0.1+\vsdoubleshift);
\filldraw (\fnresultcvOOO*\vsfactor,0.25+\vsdoubleshift) circle (2pt);

\node[text width = 15mm] at (-1,1.2+\vsdoubleshift) {$p=\pone$ $\correlations=\kappathree$};

\node at (3,2.15+\vsdoubleshift) {\avi\ false positives};
\node at (3,1.65+\vsdoubleshift) {Cross-Validation false positives};
\node at (3,0.75+\vsdoubleshift) {\avi\ false negatives};
\node at (3,0.25+\vsdoubleshift) {Cross-Validation false negatives};

\path [\coloravi,fill=\coloravi, line width = 0pt] (0,1.9+\vsshift) rectangle (\fpresultaviOO*\vsfactor,2.4+\vsshift);
\draw (\fpresultaviOO*\vsfactor-\fpresultaviEE*\vsfactor,2.15+\vsshift) -| (\fpresultaviOO*\vsfactor+\fpresultaviEE*\vsfactor,2.15+\vsshift);
\draw (\fpresultaviOO*\vsfactor-\fpresultaviEE*\vsfactor, 2.3+\vsshift) -- (\fpresultaviOO*\vsfactor-\fpresultaviEE*\vsfactor, 2+\vsshift);
\draw (\fpresultaviOO*\vsfactor+\fpresultaviEE*\vsfactor, 2.3+\vsshift) -- (\fpresultaviOO*\vsfactor+\fpresultaviEE*\vsfactor, 2+\vsshift);
\filldraw (\fpresultaviOO*\vsfactor,2.15+\vsshift) circle (2pt);
\path [\colorcv,fill=\colorcv, line width = 0pt] (0,1.4+\vsshift) rectangle (\fpresultcvOO*\vsfactor,1.9+\vsshift);
\draw (\fpresultcvOO*\vsfactor-\fpresultcvEE*\vsfactor,1.65+\vsshift) -| (\fpresultcvOO*\vsfactor+\fpresultcvEE*\vsfactor,1.65+\vsshift);
\draw (\fpresultcvOO*\vsfactor-\fpresultcvEE*\vsfactor, 1.8+\vsshift) -- (\fpresultcvOO*\vsfactor-\fpresultcvEE*\vsfactor, 1.5+\vsshift);
\draw (\fpresultcvOO*\vsfactor+\fpresultcvEE*\vsfactor, 1.8+\vsshift) -- (\fpresultcvOO*\vsfactor+\fpresultcvEE*\vsfactor, 1.5+\vsshift);
\filldraw (\fpresultcvOO*\vsfactor,1.65+\vsshift) circle (2pt);

\path [\coloravi,fill=\coloravi, line width = 0pt] (0,0.5+\vsshift) rectangle (\fnresultaviOO*\vsfactor,1+\vsshift);
\draw (\fnresultaviOO*\vsfactor-\fnresultaviEE*\vsfactor,0.75+\vsshift) -| (\fnresultaviOO*\vsfactor+\fnresultaviEE*\vsfactor,0.75+\vsshift);
\draw (\fnresultaviOO*\vsfactor-\fnresultaviEE*\vsfactor, 0.9+\vsshift) -- (\fnresultaviOO*\vsfactor-\fnresultaviEE*\vsfactor, 0.6+\vsshift);
\draw (\fnresultaviOO*\vsfactor+\fnresultaviEE*\vsfactor, 0.9+\vsshift) -- (\fnresultaviOO*\vsfactor+\fnresultaviEE*\vsfactor, 0.60+\vsshift);
\filldraw (\fnresultaviOO*\vsfactor,0.75+\vsshift) circle (2pt);
\path [\colorcv,fill=\colorcv, line width = 0pt] (0,-0.0+\vsshift) rectangle (\fnresultcvOO*\vsfactor,0.5+\vsshift);
\draw (\fnresultcvOO*\vsfactor-\fnresultcvEE*\vsfactor,0.25+\vsshift) -| (\fnresultcvOO*\vsfactor+\fnresultcvEE*\vsfactor,0.25+\vsshift);
\draw (\fnresultcvOO*\vsfactor-\fnresultcvEE*\vsfactor, 0.4+\vsshift) -- (\fnresultcvOO*\vsfactor-\fnresultcvEE*\vsfactor, 0.1+\vsshift);
\draw (\fnresultcvOO*\vsfactor+\fnresultcvEE*\vsfactor, 0.4+\vsshift) -- (\fnresultcvOO*\vsfactor+\fnresultcvEE*\vsfactor, 0.1+\vsshift);
\filldraw (\fnresultcvOO*\vsfactor,0.25+\vsshift) circle (2pt);

\node[text width = 15mm] at (-1,1.2+\vsshift) {$p=\pone$ $\correlations=\kappatwo$};

\node at (3,2.15+\vsshift) {\avi\ false positives};
\node at (3,1.65+\vsshift) {Cross-Validation false positives};
\node at (3,0.75+\vsshift) {\avi\ false negatives};
\node at (3,0.25+\vsshift) {Cross-Validation false negatives};

\path [\coloravi,fill=\coloravi, line width = 0pt] (0,1.9+\vszeroshift) rectangle (\fpresultaviO*\vsfactor,2.4+\vszeroshift);
\draw (\fpresultaviO*\vsfactor-\fpresultaviE*\vsfactor,2.15+\vszeroshift) -| (\fpresultaviO*\vsfactor+\fpresultaviE*\vsfactor,2.15+\vszeroshift);
\draw (\fpresultaviO*\vsfactor-\fpresultaviE*\vsfactor, 2.3+\vszeroshift) -- (\fpresultaviO*\vsfactor-\fpresultaviE*\vsfactor, 2+\vszeroshift);
\draw (\fpresultaviO*\vsfactor+\fpresultaviE*\vsfactor, 2.3+\vszeroshift) -- (\fpresultaviO*\vsfactor+\fpresultaviE*\vsfactor, 2+\vszeroshift);
\filldraw (\fpresultaviO*\vsfactor,2.15+\vszeroshift) circle (2pt);
\path [\colorcv,fill=\colorcv, line width = 0pt] (0,1.4+\vszeroshift) rectangle (\fpresultcvO*\vsfactor,1.9+\vszeroshift);
\draw (\fpresultcvO*\vsfactor-\fpresultcvE*\vsfactor,1.65+\vszeroshift) -| (\fpresultcvO*\vsfactor+\fpresultcvE*\vsfactor,1.65+\vszeroshift);
\draw (\fpresultcvO*\vsfactor-\fpresultcvE*\vsfactor, 1.8+\vszeroshift) -- (\fpresultcvO*\vsfactor-\fpresultcvE*\vsfactor, 1.5+\vszeroshift);
\draw (\fpresultcvO*\vsfactor+\fpresultcvE*\vsfactor, 1.8+\vszeroshift) -- (\fpresultcvO*\vsfactor+\fpresultcvE*\vsfactor, 1.5+\vszeroshift);
\filldraw (\fpresultcvO*\vsfactor,1.65+\vszeroshift) circle (2pt);

\path [\coloravi,fill=\coloravi, line width = 0pt] (0,0.5+\vszeroshift) rectangle (\fnresultaviO*\vsfactor,1+\vszeroshift);
\draw (\fnresultaviO*\vsfactor-\fnresultaviE*\vsfactor,0.75+\vszeroshift) -| (\fnresultaviO*\vsfactor+\fnresultaviE*\vsfactor,0.75+\vszeroshift);
\draw (\fnresultaviO*\vsfactor-\fnresultaviE*\vsfactor, 0.9+\vszeroshift) -- (\fnresultaviO*\vsfactor-\fnresultaviE*\vsfactor, 0.6+\vszeroshift);
\draw (\fnresultaviO*\vsfactor+\fnresultaviE*\vsfactor, 0.9+\vszeroshift) -- (\fnresultaviO*\vsfactor+\fnresultaviE*\vsfactor, 0.60+\vszeroshift);
\filldraw (\fnresultaviO*\vsfactor,0.75+\vszeroshift) circle (2pt);
\path [\colorcv,fill=\colorcv, line width = 0pt] (0,-0.0+\vszeroshift) rectangle (\fnresultcvO*\vsfactor,0.5+\vszeroshift);
\draw (\fnresultcvO*\vsfactor-\fnresultcvE*\vsfactor,0.25+\vszeroshift) -| (\fnresultcvO*\vsfactor+\fnresultcvE*\vsfactor,0.25+\vszeroshift);
\draw (\fnresultcvO*\vsfactor-\fnresultcvE*\vsfactor, 0.4+\vszeroshift) -- (\fnresultcvO*\vsfactor-\fnresultcvE*\vsfactor, 0.1+\vszeroshift);
\draw (\fnresultcvO*\vsfactor+\fnresultcvE*\vsfactor, 0.4+\vszeroshift) -- (\fnresultcvO*\vsfactor+\fnresultcvE*\vsfactor, 0.1+\vszeroshift);
\filldraw (\fnresultcvO*\vsfactor,0.25+\vszeroshift) circle (2pt);

\node[text width = 15mm] at (-1,1.2+\vszeroshift) {$p=\pone$ $\correlations=\kappaone$};

\node at (3,2.15+\vszeroshift) {\avi\ false positives};
\node at (3,1.65+\vszeroshift) {Cross-Validation false positives};
\node at (3,0.75+\vszeroshift) {\avi\ false negatives};
\node at (3,0.25+\vszeroshift) {Cross-Validation false negatives};

\draw[line width = 1pt, black!80, dotted] (20*\vsfactor,-5.8) --(20*\vsfactor,5.3);
\draw[line width = 1pt, black!80, dotted] (40*\vsfactor,-5.8) --(40*\vsfactor,5.3);
\draw[line width = 1pt, black!80, dotted] (60*\vsfactor,-5.8) --(60*\vsfactor,5.3);
\draw[line width = 3pt] (-0,-5.8) --(0,5.3); 
\draw[line width = 3pt, ->] (-0.3,-5.5) --(14,-5.5); 
\node at (14,-6.2) {Variable selection}; 
\node at (14,-6.6) {error}; 
\node at (0,-6.2) {$0$};
\node at (20*\vsfactor,-6.2) {$20$};
\node at (40*\vsfactor,-6.2) {$40$};
\node at (60*\vsfactor,-6.2) {$60$};
 
\end{tikzpicture}
\caption{\captionvs}
\label{MainComparison3}
\end{center}
\end{figure}


\begin{figure}
\begin{tikzpicture}
  
\path [\coloravi,fill=\coloravi, line width = 0pt] (0,1.9+\vsdoubleshift) rectangle (\SfpresultaviOOO*\vsfactor,2.4+\vsdoubleshift);
\draw (\SfpresultaviOOO*\vsfactor-\SfpresultaviEEE*\vsfactor,2.15+\vsdoubleshift) -| (\SfpresultaviOOO*\vsfactor+\SfpresultaviEEE*\vsfactor,2.15+\vsdoubleshift);
\draw (\SfpresultaviOOO*\vsfactor-\SfpresultaviEEE*\vsfactor, 2.3+\vsdoubleshift) -- (\SfpresultaviOOO*\vsfactor-\SfpresultaviEEE*\vsfactor, 2+\vsdoubleshift);
\draw (\SfpresultaviOOO*\vsfactor+\SfpresultaviEEE*\vsfactor, 2.3+\vsdoubleshift) -- (\SfpresultaviOOO*\vsfactor+\SfpresultaviEEE*\vsfactor, 2+\vsdoubleshift);
\filldraw (\SfpresultaviOOO*\vsfactor,2.15+\vsdoubleshift) circle (2pt);
\path [\colorcv,fill=\colorcv, line width = 0pt] (0,1.4+\vsdoubleshift) rectangle (\SfpresultcvOOO*\vsfactor,1.9+\vsdoubleshift);
\draw (\SfpresultcvOOO*\vsfactor-\SfpresultcvEEE*\vsfactor,1.65+\vsdoubleshift) -| (\SfpresultcvOOO*\vsfactor+\SfpresultcvEEE*\vsfactor,1.65+\vsdoubleshift);
\draw (\SfpresultcvOOO*\vsfactor-\SfpresultcvEEE*\vsfactor, 1.8+\vsdoubleshift) -- (\SfpresultcvOOO*\vsfactor-\SfpresultcvEEE*\vsfactor, 1.5+\vsdoubleshift);
\draw (\SfpresultcvOOO*\vsfactor+\SfpresultcvEEE*\vsfactor, 1.8+\vsdoubleshift) -- (\SfpresultcvOOO*\vsfactor+\SfpresultcvEEE*\vsfactor, 1.5+\vsdoubleshift);
\filldraw (\SfpresultcvOOO*\vsfactor,1.65+\vsdoubleshift) circle (2pt);

\path [\coloravi,fill=\coloravi, line width = 0pt] (0,0.5+\vsdoubleshift) rectangle (\SfnresultaviOOO*\vsfactor,1+\vsdoubleshift);
\draw (\SfnresultaviOOO*\vsfactor-\SfnresultaviEEE*\vsfactor,0.75+\vsdoubleshift) -| (\SfnresultaviOOO*\vsfactor+\SfnresultaviEEE*\vsfactor,0.75+\vsdoubleshift);
\draw (\SfnresultaviOOO*\vsfactor-\SfnresultaviEEE*\vsfactor, 0.9+\vsdoubleshift) -- (\SfnresultaviOOO*\vsfactor-\SfnresultaviEEE*\vsfactor, 0.6+\vsdoubleshift);
\draw (\SfnresultaviOOO*\vsfactor+\SfnresultaviEEE*\vsfactor, 0.9+\vsdoubleshift) -- (\SfnresultaviOOO*\vsfactor+\SfnresultaviEEE*\vsfactor, 0.60+\vsdoubleshift);
\filldraw (\SfnresultaviOOO*\vsfactor,0.75+\vsdoubleshift) circle (2pt);
\path [\colorcv,fill=\colorcv, line width = 0pt] (0,-0.0+\vsdoubleshift) rectangle (\SfnresultcvOOO*\vsfactor,0.5+\vsdoubleshift);
\draw (\SfnresultcvOOO*\vsfactor-\SfnresultcvEEE*\vsfactor,0.25+\vsdoubleshift) -| (\SfnresultcvOOO*\vsfactor+\SfnresultcvEEE*\vsfactor,0.25+\vsdoubleshift);
\draw (\SfnresultcvOOO*\vsfactor-\SfnresultcvEEE*\vsfactor, 0.4+\vsdoubleshift) -- (\SfnresultcvOOO*\vsfactor-\SfnresultcvEEE*\vsfactor, 0.1+\vsdoubleshift);
\draw (\SfnresultcvOOO*\vsfactor+\SfnresultcvEEE*\vsfactor, 0.4+\vsdoubleshift) -- (\SfnresultcvOOO*\vsfactor+\SfnresultcvEEE*\vsfactor, 0.1+\vsdoubleshift);
\filldraw (\SfnresultcvOOO*\vsfactor,0.25+\vsdoubleshift) circle (2pt);

\node[text width = 15mm] at (-1,1.2+\vsdoubleshift) {$p=\ptwo$ $\correlations=\kappathree$};

\node at (3,2.15+\vsdoubleshift) {\avi\ false positives};
\node at (3,1.65+\vsdoubleshift) {Cross-Validation false positives};
\node at (3,0.75+\vsdoubleshift) {\avi\ false negatives};
\node at (3,0.25+\vsdoubleshift) {Cross-Validation false negatives};

\path [\coloravi,fill=\coloravi, line width = 0pt] (0,1.9+\vsshift) rectangle (\SfpresultaviOO*\vsfactor,2.4+\vsshift);
\draw (\SfpresultaviOO*\vsfactor-\SfpresultaviEE*\vsfactor,2.15+\vsshift) -| (\SfpresultaviOO*\vsfactor+\SfpresultaviEE*\vsfactor,2.15+\vsshift);
\draw (\SfpresultaviOO*\vsfactor-\SfpresultaviEE*\vsfactor, 2.3+\vsshift) -- (\SfpresultaviOO*\vsfactor-\SfpresultaviEE*\vsfactor, 2+\vsshift);
\draw (\SfpresultaviOO*\vsfactor+\SfpresultaviEE*\vsfactor, 2.3+\vsshift) -- (\SfpresultaviOO*\vsfactor+\SfpresultaviEE*\vsfactor, 2+\vsshift);
\filldraw (\SfpresultaviOO*\vsfactor,2.15+\vsshift) circle (2pt);
\path [\colorcv,fill=\colorcv, line width = 0pt] (0,1.4+\vsshift) rectangle (\SfpresultcvOO*\vsfactor,1.9+\vsshift);
\draw (\SfpresultcvOO*\vsfactor-\SfpresultcvEE*\vsfactor,1.65+\vsshift) -| (\SfpresultcvOO*\vsfactor+\SfpresultcvEE*\vsfactor,1.65+\vsshift);
\draw (\SfpresultcvOO*\vsfactor-\SfpresultcvEE*\vsfactor, 1.8+\vsshift) -- (\SfpresultcvOO*\vsfactor-\SfpresultcvEE*\vsfactor, 1.5+\vsshift);
\draw (\SfpresultcvOO*\vsfactor+\SfpresultcvEE*\vsfactor, 1.8+\vsshift) -- (\SfpresultcvOO*\vsfactor+\SfpresultcvEE*\vsfactor, 1.5+\vsshift);
\filldraw (\SfpresultcvOO*\vsfactor,1.65+\vsshift) circle (2pt);

\path [\coloravi,fill=\coloravi, line width = 0pt] (0,0.5+\vsshift) rectangle (\SfnresultaviOO*\vsfactor,1+\vsshift);
\draw (\SfnresultaviOO*\vsfactor-\SfnresultaviEE*\vsfactor,0.75+\vsshift) -| (\SfnresultaviOO*\vsfactor+\SfnresultaviEE*\vsfactor,0.75+\vsshift);
\draw (\SfnresultaviOO*\vsfactor-\SfnresultaviEE*\vsfactor, 0.9+\vsshift) -- (\SfnresultaviOO*\vsfactor-\SfnresultaviEE*\vsfactor, 0.6+\vsshift);
\draw (\SfnresultaviOO*\vsfactor+\SfnresultaviEE*\vsfactor, 0.9+\vsshift) -- (\SfnresultaviOO*\vsfactor+\SfnresultaviEE*\vsfactor, 0.60+\vsshift);
\filldraw (\SfnresultaviOO*\vsfactor,0.75+\vsshift) circle (2pt);
\path [\colorcv,fill=\colorcv, line width = 0pt] (0,-0.0+\vsshift) rectangle (\SfnresultcvOO*\vsfactor,0.5+\vsshift);
\draw (\SfnresultcvOO*\vsfactor-\SfnresultcvEE*\vsfactor,0.25+\vsshift) -| (\SfnresultcvOO*\vsfactor+\SfnresultcvEE*\vsfactor,0.25+\vsshift);
\draw (\SfnresultcvOO*\vsfactor-\SfnresultcvEE*\vsfactor, 0.4+\vsshift) -- (\SfnresultcvOO*\vsfactor-\SfnresultcvEE*\vsfactor, 0.1+\vsshift);
\draw (\SfnresultcvOO*\vsfactor+\SfnresultcvEE*\vsfactor, 0.4+\vsshift) -- (\SfnresultcvOO*\vsfactor+\SfnresultcvEE*\vsfactor, 0.1+\vsshift);
\filldraw (\SfnresultcvOO*\vsfactor,0.25+\vsshift) circle (2pt);

\node[text width = 15mm] at (-1,1.2+\vsshift) {$p=\ptwo$ $\correlations=\kappatwo$};

\node at (3,2.15+\vsshift) {\avi\ false positives};
\node at (3,1.65+\vsshift) {Cross-Validation false positives};
\node at (3,0.75+\vsshift) {\avi\ false negatives};
\node at (3,0.25+\vsshift) {Cross-Validation false negatives};

\path [\coloravi,fill=\coloravi, line width = 0pt] (0,1.9+\vszeroshift) rectangle (\SfpresultaviO*\vsfactor,2.4+\vszeroshift);
\draw (\SfpresultaviO*\vsfactor-\SfpresultaviE*\vsfactor,2.15+\vszeroshift) -| (\SfpresultaviO*\vsfactor+\SfpresultaviE*\vsfactor,2.15+\vszeroshift);
\draw (\SfpresultaviO*\vsfactor-\SfpresultaviE*\vsfactor, 2.3+\vszeroshift) -- (\SfpresultaviO*\vsfactor-\SfpresultaviE*\vsfactor, 2+\vszeroshift);
\draw (\SfpresultaviO*\vsfactor+\SfpresultaviE*\vsfactor, 2.3+\vszeroshift) -- (\SfpresultaviO*\vsfactor+\SfpresultaviE*\vsfactor, 2+\vszeroshift);
\filldraw (\SfpresultaviO*\vsfactor,2.15+\vszeroshift) circle (2pt);
\path [\colorcv,fill=\colorcv, line width = 0pt] (0,1.4+\vszeroshift) rectangle (\SfpresultcvO*\vsfactor,1.9+\vszeroshift);
\draw (\SfpresultcvO*\vsfactor-\SfpresultcvE*\vsfactor,1.65+\vszeroshift) -| (\SfpresultcvO*\vsfactor+\SfpresultcvE*\vsfactor,1.65+\vszeroshift);
\draw (\SfpresultcvO*\vsfactor-\SfpresultcvE*\vsfactor, 1.8+\vszeroshift) -- (\SfpresultcvO*\vsfactor-\SfpresultcvE*\vsfactor, 1.5+\vszeroshift);
\draw (\SfpresultcvO*\vsfactor+\SfpresultcvE*\vsfactor, 1.8+\vszeroshift) -- (\SfpresultcvO*\vsfactor+\SfpresultcvE*\vsfactor, 1.5+\vszeroshift);
\filldraw (\SfpresultcvO*\vsfactor,1.65+\vszeroshift) circle (2pt);

\path [\coloravi,fill=\coloravi, line width = 0pt] (0,0.5+\vszeroshift) rectangle (\SfnresultaviO*\vsfactor,1+\vszeroshift);
\draw (\SfnresultaviO*\vsfactor-\SfnresultaviE*\vsfactor,0.75+\vszeroshift) -| (\SfnresultaviO*\vsfactor+\SfnresultaviE*\vsfactor,0.75+\vszeroshift);
\draw (\SfnresultaviO*\vsfactor-\SfnresultaviE*\vsfactor, 0.9+\vszeroshift) -- (\SfnresultaviO*\vsfactor-\SfnresultaviE*\vsfactor, 0.6+\vszeroshift);
\draw (\SfnresultaviO*\vsfactor+\SfnresultaviE*\vsfactor, 0.9+\vszeroshift) -- (\SfnresultaviO*\vsfactor+\SfnresultaviE*\vsfactor, 0.60+\vszeroshift);
\filldraw (\SfnresultaviO*\vsfactor,0.75+\vszeroshift) circle (2pt);
\path [\colorcv,fill=\colorcv, line width = 0pt] (0,-0.0+\vszeroshift) rectangle (\SfnresultcvO*\vsfactor,0.5+\vszeroshift);
\draw (\SfnresultcvO*\vsfactor-\SfnresultcvE*\vsfactor,0.25+\vszeroshift) -| (\SfnresultcvO*\vsfactor+\SfnresultcvE*\vsfactor,0.25+\vszeroshift);
\draw (\SfnresultcvO*\vsfactor-\SfnresultcvE*\vsfactor, 0.4+\vszeroshift) -- (\SfnresultcvO*\vsfactor-\SfnresultcvE*\vsfactor, 0.1+\vszeroshift);
\draw (\SfnresultcvO*\vsfactor+\SfnresultcvE*\vsfactor, 0.4+\vszeroshift) -- (\SfnresultcvO*\vsfactor+\SfnresultcvE*\vsfactor, 0.1+\vszeroshift);
\filldraw (\SfnresultcvO*\vsfactor,0.25+\vszeroshift) circle (2pt);

\node[text width = 15mm] at (-1,1.2+\vszeroshift) {$p=\ptwo$ $\correlations=\kappaone$};

\node at (3,2.15+\vszeroshift) {\avi\ false positives};
\node at (3,1.65+\vszeroshift) {Cross-Validation false positives};
\node at (3,0.75+\vszeroshift) {\avi\ false negatives};
\node at (3,0.25+\vszeroshift) {Cross-Validation false negatives};

\draw[line width = 1pt, black!80, dotted] (20*\vsfactor,-5.8) --(20*\vsfactor,5.3);
\draw[line width = 1pt, black!80, dotted] (40*\vsfactor,-5.8) --(40*\vsfactor,5.3);
\draw[line width = 1pt, black!80, dotted] (60*\vsfactor,-5.8) --(60*\vsfactor,5.3);
\draw[line width = 3pt] (-0,-5.8) --(0,5.3); 
\draw[line width = 3pt, ->] (-0.3,-5.5) --(14,-5.5); 
\node at (14,-6.2) {Variable selection}; 
\node at (14,-6.6) {error}; 
\node at (0,-6.2) {$0$};
\node at (20*\vsfactor,-6.2) {$20$};
\node at (40*\vsfactor,-6.2) {$40$};
\node at (60*\vsfactor,-6.2) {$60$};

\end{tikzpicture}
\caption{\captionvst}
\label{MainComparison4}
\end{figure}

We compare the sup-norm and variable selection performance of the
following three procedures:
\begin{itemize}
\item[-] \nameplotoracle: Lasso with the tuning parameter that minimizes the $\ell_\infty$ loss ({\it this tuning parameter is unknown in practice});
 \item[-] \avi: Lasso with \avi~and $\KEYCONTWO=0.75$;
\item[-] Cross-Validation: Lasso with $10$-fold Cross-Validation.
\end{itemize}
\revision{Our choice $\KEYCONTWO=0.75$ is motivated by a theorem due
  to~\cite{Lounici08} in the regime $\alpha\to\infty$; see
  Remark~\ref{motivationC} for details.}

\emph{Sup-norm error: } In Figures~\ref{MainComparison1}
and~\ref{MainComparison2}, we compare the $\ell_\infty$ error of the
four procedures. We observe that \avi\ outperforms Cross-Validation
for most settings under consideration.  We also mention that the same
conclusions can be drawn if the normal distribution for the noise is
replaced by other, possibly heavy-tailed distributions (for
conciseness, we do not show the outputs).

\emph{Variable selection:} In Figures~\ref{MainComparison3}
and~\ref{MainComparison4}, we compare the variable selection
performance of \avi\ and Cross-Validation. More specifically, we
compare the number of false positives $|\{j:\bt_j=
0,(\widehat\beta_\lambda)_j\neq 0\}|$ and the number of false
negatives $|\{j:\bt_j\neq 0,(\widehat\beta_\lambda)_j= 0\}|.$ In
contrast to Cross-Validation, \avi~allows for a safe threshold of size
$3\KEYCONTWO\lamhat$ (recall the discussion after
Theorem~\ref{ThmAvi}). Therefore, we report the results of Lasso with
\avi\ and an additional threshold of size $3\KEYCONTWO\lamhat$ applied
to each component (that is, we consider the vector with entries
$(\widehat\beta_\lambda)_j\1\large\{|(\widehat\beta_\lambda)_j|\geq
3\KEYCONTWO\lamhat\large\}$ ), and we report the results of Lasso with
Cross-Validation (without threshold). We observe that, as compared to
Cross-Validation, \avi\ with subsequent thresholding can lead to a
considerably smaller number of false positives, while keeping the
number of false negatives on a low level. Note that one could perform
a similar thresholding of the Cross-Validation solution, but unlike
for~\avi, there is no theory to guide the choice of the
threshold. This problem also applies to other standard calibration
schemes.

\emph{Computational complexity:} Cross-Validation with $10$ folds
requires the computation $10$ Lasso paths, while \avi~requires the
computation of only one Lasso path - or even less. \avi~is therefore
about $10$ times more efficient than $10$-fold Cross-Validation.

\revision{Let us conclude with remarks on the scope of the
  simulations. First, many methods have been proposed for tuning the
  regularization parameter in the Lasso, including Cross-Validation,
  BIC and AIC-type criteria, Stability
  Selection~\citep{Meinshausen10},
  LinSelect~\citep{Baraud14,Giraud12}, permutation
  approaches~\citep{Sabourin15}, and many more. On top of that, there
  are many modifications and extensions of the Lasso itself, including
  BoLasso~\citep{Bach08}, Square-Root/Scaled
  Lasso~\citep{Antoniadis10,Belloni11,Stadler10,ScaledLasso11},
  SCAD~\citep{Fan_Li01}, MCP~\citep{Zhang10}, and others. Detailed
  comparisons among the selection schemes and the methods can be found
  in the cited papers. We also refer to \cite{Leeb08} for theoretical
  insights about limitations of the methods.  

In our simulations, we instead focus on the Lasso and, since we are
not aware of guarantees similar to ours for any selection scheme, we
compare to the most popular and most extensively studied selection
scheme, Cross-Validation. This comparison shows that, beyond its
theoretical properties and the easy and efficient implementation,
\avi\ is also a competitor in numerical experiments.}


\subsection{Riboflavin Production in {\it B. subtilis}}
We now consider variable selection for a data set that describes the
production of riboflavin (vitamin B${}_2$) in {\it B.~subtilis} ({\it
  Bacillus subtilis}), see~\citep{Buhlmann:2014}. The data set
comprises the expressions of $p = 4088$ genes and the corresponding
riboflavin production rates for $n = 71$ strains of {\it
  B.~subtilis}. We apply \avi\ and then impose the
threshold~$3\KEYCONTWO\lamhat$.

The resulting genes and the corresponding parameter values are given
in the first column of Table~\ref{fig:ribo}. We see that these results
commensurate with the results from previous approaches based on
Stability Selection~\citep{Buhlmann:2014} and TREX~\citep{Lederer:14},
which are given in the third and fourth column.

\begin{table}
\begin{center}
\begin{scriptsize}
\begin{tabular}[h]{c c c c c}
\multicolumn{2}{c}{\avi}\vspace{2mm}&~~~~Stability
Selection~~~~&~~~B-TREX\hspace{-0.3mm}~~~~\\ YXLD\underline{~}at&\hspace{-3mm}\text-0.405&YXLD\underline{~}at&YXLD\underline{~}at\\ YOAB\underline{~}at
&\hspace{-3mm}\text-0.420&YOAB\underline{~}at&YOAB\underline{~}at\\ YEBC\underline{~}at&\hspace{-3mm}\text-0.146&LYSC\underline{~}at&YXLE\underline{~}at\\ ARGF\underline{~}at&\hspace{-3mm}\text-0.313\\ XHLB\underline{~}at&\hspace{-2.2mm}0.278
\end{tabular}
\end{scriptsize}
\caption{Variable selection results for the riboflavin data set. The
  first column depicts the genes and the corresponding parameter
  values yielded by \avi. The second and third column depict the genes
  returned by approaches based on Stability Selection and TREX.}
\label{fig:ribo}
\end{center}
\end{table}



\section{Conclusions}
\label{conclusions}
\jaj{We have introduced a novel method for sup-norm calibration, known
  as \avi, that is equipped with finite sample guarantees for
  estimation in $\ell_\infty$-loss and for variable
  selection. Moreover, we have shown that \avi\ allows for simple and
  fast implementations.  These properties make \avi\ a competitive
  algorithm, as standard methods such as Cross-Validation are
  computationally more demanding and lack non-asymptotic guarantees.}

In order to bring sharp focus to the issue, we have focused this paper
exclusively on the calibration of the Lasso. However, we suspect that
the methods and techniques developed here could be more generally
applicable, for instance to problems with nonconvex penalties (e.g.,
SCAD, MCP). In particular, the paper~\citep{Loh14} provides guarantees
for $\ell_\infty$-recovery using such nonconvex methods, which could
be combined with our results. Another interesting direction for future
work is the use of our methods for more general decomposable penalty
functions~\citep{NegRavWaiYu12}, including the nuclear norm that is
often used in matrix estimation.

\revision{We also stress that our goals are $\ell_\infty$-estimation
  and variable selection, which are feasible only under strict
  conditions on the design matrix. Other objectives, including
  prediction and $\ell_2$-estimation, can typically be achieved under
  less stringent conditions. However, the corresponding oracle
  inequalities contain quantities (such as the sparsity level) that
  are typically unknown in practice. Adaptations of our method to
  objectives beyond the ones considered here thus need further
  investigation.  We refer to \citep{Chetelat14} for ideas in this
  direction. However, there might be no approach that is uniformly
  optimal for all objectives, see also the
  papers~\citep{Yang05,Zhao06}.}

\revision{Finally, as pointed out by one of the reviewers, another
  field for further study is model misspecification. It would be
  interesting to see how robust the Lasso with the \avi\ scheme is
  with respect to, for example, non-linearities in the model.}


~\\
\acks{We thank Sara van de Geer and S\'ebastien Loustau for the
  inspiring discussions. We also thank the reviewers for the careful
  reading of the manuscript and the insightful comments.  This work
  was partially supported by NSF Grant DMS-1107000, and Air Force
  Office of Scientific Research AFOSR-FA9550-14-1-0016 to MJW.}


\appendix

\section{Proof of Theorem 1}
\label{AppThmAvi}

Define the event $\GoodEventStar \defn \left \{ \frac{\|\Xmat^\top
  \varepsilon\|_\infty}{\numobs} \leq \frac{\ot}{4} \right\}$ and note
that $\mpr[ \GoodEventStar] \geq 1 - \delta$ by our definition of the
oracle tuning parameter in \eqref{EqnOracleParameter}.  Thus, it
suffices to show that the two bounds hold conditioned on the event
$\GoodEventStar$.

\paragraph{Bound on $\lamhat$:} To show that $\lamhat \leq \ot$, we
proceed by proof by contradiction.  If $\lamhat > \ot$, then the
definition of the \avi~method implies that there must exist two tuning
parameters $\lambda',\lambda''\geq \ot $ such that
\begin{align}
\label{EqnBeijingBreakfast}
\|\betahat_{\lambda'}-\betahat_{\lambda''}\|_\infty> \KEYCONTWO \,
(\lambda'+\lambda'').
\end{align}
However, since $\GoodEvent{\lambda'}$ and $\GoodEvent{\lambda''}$ are
both subsets of $\GoodEventStar$, Assumption~$\ell_\infty(\KEYCON)$
implies that we must have the simultaneous inequalities
\mbox{$\|\betahat_{\lambda'} - \bt\|_\infty \leq \KEYCON \lambda'$}
and \mbox{$\|\betahat_{\lambda''} - \bt \|_\infty \leq \KEYCON
  \lambda''$.}  Combined with the triangle inequality, we find that
\begin{align*}
\|\betahat_{\lambda'}-\betahat_{\lambda''}\|_\infty\leq
\|\betahat_{\lambda'}-\bt\|_\infty+\|\bt-
\betahat_{\lambda''}\|_\infty \; \leq \; \KEYCON \, (\lambda' +
\lambda'').
\end{align*}
Since $\KEYCONTWO\geq \KEYCON$, this upper bound contradicts our earlier
conclusion~\eqref{EqnBeijingBreakfast} and, therefore, yields the
desired claim.


\paragraph{Bound on the sup-norm error:}

On the event $\GoodEventStar$, we have $\lamhat \leq \ot$, and so the
\avi~definition implies that
\begin{align*}
\|\betahat_{\lamhat} - \betahat_{\ot}\|_\infty & \leq \KEYCONTWO \, \big
( \lamhat + \ot \big) \, \leq \, 2 \KEYCONTWO \ot.
\end{align*}
Combined with the triangle inequality, we find that
\begin{align*}
\|\betahat_{\lamhat}-\bt\|_\infty \leq
\|\betahat_{\lamhat}-\betahat_{\ot}\|_\infty +
\|\betahat_{\ot}-\bt\|_\infty \leq 2 \KEYCONTWO \ot + \|\betahat_{\ot} -
\bt\|_\infty.
\end{align*}
Finally, under $\GoodEventStar$ and $\KEYCONTWO\geq\KEYCON$,  Assumption~$\ell_\infty(\KEYCON)$
implies that $\|\betahat_{\ot} - \bt\|_\infty \leq \KEYCON \ot\leq \KEYCONTWO \ot$, and
combining the pieces completes the proof.{\hfill$\blacksquare$}


\section{Remaining Proofs for Section 2}
\label{AppSecCompatibility}

In this appendix, we provide the proofs of
Lemmas~\ref{LemGammaRE},~\ref{LemDiagonalDominant}, and~\ref{multlem}.


\subsection{Proof of Lemma~\ref{LemGammaRE}}\label{ProofLemma1}

By the first-order stationarity conditions for an optimum, the Lasso
solution $\betalasso$ must satisfy the stationary condition
$\frac{1}{\numobs} \Xmat^\top \big( \Xmat \betalasso - Y \big) +
\lambda \zhat = 0$, where $\zhat \in \real^\pdim$ belongs to the
sub-differential of the $\ell_1$-norm at $\betalasso$.  Since $Y =
\Xmat \betastar + \varepsilon$, we find that
\begin{align*}
\SamCov \big(\betalasso - \betastar \big) & = - \lambda \zhat +
\frac{\Xmat^\top \varepsilon}{\numobs}.
\end{align*}
Taking the $\ell_\infty$-norm of both sides and applying the triangle
inequality yields
\begin{align*}
\|\SamCov \big(\betalasso - \betastar \big)\|_\infty \leq \lambda
\|\zhat\|_\infty + \left \|\frac{\Xmat^\top \varepsilon}{\numobs} \right
\|_\infty \leq \lambda + \frac{\lambda}{4} \; = \; \frac{5}{4}
\lambda,
\end{align*}
using the bound from event $\GoodEvent{\lambda}$, and the fact that
$\|\zhat\|_\infty \leq 1$, by definition of the
$\ell_1$-sub-differential.  As noted previously, under the event
$\GoodEvent{\lambda}$, the error vector $\DelHat = \betalasso -
\betastar$ belongs to the cone $\Cone(\Ssettrue)$ in~\eqref{ConeDefn},
so that the $\gamma$-RE condition can be applied so as to obtain the
lower bou \mbox{$\|\SamCov (\betalasso - \betastar)\|_\infty \geq
  \gamma \|\betalasso - \betastar\|_\infty$.}  Combining the pieces
concludes the proof.{\hfill$\blacksquare$}


\subsection{Proof of Lemma~\ref{LemDiagonalDominant}}\label{ProofLemma2}

\newcommand{\cl}{\operatorname{cl}}
\newcommand{\convex}{\operatorname{conv}}
\newcommand{\Ball}{\ensuremath{\mathbb{B}}}
\newcommand{\Ssettruenorm}{{|\Ssettrue|}}

Since $\Delta \in \Cone(\Ssettrue)$, we have 
\begin{equation*}
  \|\Delta\|_1^2 \leq 9 \|\Delta_{\Ssettrue}\|_1^2 \leq 9
  \Ssettruenorm \|\Delta_{\Ssettrue}\|_2^2\leq 9 \Ssettruenorm
  \|\Delta\|_2^2 \leq 9\Ssettruenorm \|\Delta\|_1 \|\Delta\|_\infty,
\end{equation*}
which implies $\|\Delta\|_1 \leq 9 \Ssettruenorm
\|\Delta\|_\infty$. In view of Lemma~\ref{LemGammaRE}, it thus
suffices to prove the lower bound
\begin{align}
\label{EqnIntermediate}
\|\SamCov \Delta\|_\infty \geq (1 - \nu) \|\Delta\|_\infty \qquad
\mbox{for all $\Delta \in A \defn \Ball_1(9 \Ssettruenorm) \cap
  \Ball_\infty(1)$,}
\end{align}
where we set $\Ball_d(r) \defn \{\beta\in\R^p:\|\beta\|_d\leq r\}$ for
$d\in[0,\infty]$ and $r\geq 0$. We claim that
\begin{align}
\label{EqnInclusion}
\underbrace{\Ball_1(9 \Ssettruenorm) \cap \Ball_\infty(1)}_{A}
\subseteq \underbrace{2\cl\convex \big \{ \Ball_0(9\Ssettruenorm) \cap
  \Ball_\infty(1) \}}_{B},
\end{align}
where $\cl\convex$ denotes the closed convex hull.  Taking this as
given for the moment, let us use it to prove the desired claim.  We
have
\begin{align}
\max_{\Delta \in A} \frac{\|(\SamCov - \operatorname{I}) \Delta
  \|_\infty}{\|\Delta\|_\infty} =\max_{\Delta \in A/2}
\frac{\|(\SamCov - \operatorname{I}) \Delta
  \|_\infty}{\|\Delta\|_\infty} & \leq \, \max_{\Delta \in B}
\frac{\|(\SamCov - \operatorname{I}) \Delta
  \|_\infty}{\|\Delta\|_\infty} \; \leq \max_{j \in [\pdim]}
\max_{\substack{|T| = 9 \Ssettruenorm \\ T \subset [\pdim] \backslash
    j}} \sum_{k \in T} |\SamCov_{jk}| \; \leq \; \nu
\end{align}
using the diagonal dominance~\eqref{EqnDiagonalDominant}. Combined
with the triangle inequality, the lower bound~\eqref{EqnIntermediate}
follows.

It remains to prove the inclusion~\eqref{EqnInclusion}.  Since both
$A$ and $B$ are closed and convex, it suffices to prove that
$\phi_A(\theta) \leq \phi_B(\theta)$ for all $\theta \in \real^\pdim$,
where $\phi_A(\theta) \defn \sup_{z \in A} \inprod{z}{\theta}$ and
$\phi_B(\theta) \defn \sup_{z \in B} \inprod{z}{\theta}$ are the
support functions.  For a given vector $\theta \in \real^\pdim$, let
$T$ be the subset indexing its top $9 \Ssettruenorm$ values in
absolute value.  By construction, we are guaranteed to have the bound
\mbox{$9\Ssettruenorm\|\theta_{T^c}\|_\infty \leq {\|\theta_T\|_1}$,}
and consequently
\begin{align*}
\sup_{z \in A}
\left(\inprod{z_T}{\theta_T}+\inprod{z_{T^C}}{\theta_{T^C}}\right)
\phi_A(\theta) & \leq \sup_{z \in A}
\left(\|z_T\|_\infty\|\theta_T\|_1+\|z_{T^C}\|_1\|\theta_{T^C}\|_\infty\right)
\\
& \leq \|\theta_T\|_1 + 9 \Ssettruenorm \, \|\theta_{T^c}\|_\infty \\
& \leq \; 2 \|\theta_T\|_1.
\end{align*}
On the other hand, for this same subset $T$, we have $\phi_B(\theta)
\geq \sup_{z\in B} \inprod{z_T}{\theta_T} \; = \; 2 \|\theta_T\|_1$,
which completes the proof.{\hfill$\blacksquare$}


\subsection{Proof of Lemma~\ref{multlem}}
\label{sec:proofsmultfox}

In order to prove Lemma~\ref{multlem}, we use a somewhat simplified
version of a recent result due to~\cite{vdGeer14}.  So as to simplify
notation, we first define the norms $\|a\|_j\defn|a_j|$ and $\|a\|_{-
  j}\defn\sum_{i\neq j}|a_i|$ for any vector $a$. We then have:


\begin{lemma}[\cite{vdGeer14}, Lemma 2.1]
\label{multlemhelp}
Given any tuning parameter $\lambda>0$, it holds that
\begin{equation*}
  \|\betahat_\lambda-\bt\|_j\leq
  D_j\left(\frac{\|X^\top\varepsilon\|_\infty}{n}+\frac{\sqrt{\log(p)}\|\betahat_\lambda-\bt\|_{-j}}{2\sqrt{n}\|\eta^j\|_1}+\frac{\lambda}{2}\right)
  \qquad \mbox{for all $j = 1, \ldots, \pdim$,}
\end{equation*}
where for each $j\in\otp$,
\begin{equation*}
     D_j\defn\frac{\|\eta^j\|_1}{\|X\eta^j\|_2^2/n+\ujs\|\eta^j\|_{-j}/2}.
  \end{equation*}
\end{lemma}
\noindent This result provides a specific bound for each coordinate of
Lasso. Lemma~\ref{multlem} can then readily be proven using this
result together with Theorem 6.1 from~\cite{Buhlmann11}.
~{\hfill$\blacksquare$}

 \newcommand{\factors}{4}

\newcommand{\resultbestOs}{0.75}
\newcommand{\resultbestEs}{0.2}
\newcommand{\fpresultbestOs}{0}
\newcommand{\fpresultbestEs}{0}
\newcommand{\fnresultbestOs}{0}
\newcommand{\fnresultbestEs}{0}
\newcommand{\resultaviOs}{1.82}
\newcommand{\resultaviEs}{1.43}
\newcommand{\fpresultaviOs}{0.29}
\newcommand{\fpresultaviEs}{0.64}
\newcommand{\fnresultaviOs}{2.91}
\newcommand{\fnresultaviEs}{1.14}
\newcommand{\resultcvOs}{0.88}
\newcommand{\resultcvEs}{0.25}
\newcommand{\fpresultcvOs}{22}
\newcommand{\fpresultcvEs}{13.95}
\newcommand{\fnresultcvOs}{0.99}
\newcommand{\fnresultcvEs}{1.3}

\newcommand{\resultbestOOs}{0.88}
\newcommand{\resultbestEEs}{0.23}
\newcommand{\fpresultbestOOs}{0}
\newcommand{\fpresultbestEEs}{0}
\newcommand{\fnresultbestOOs}{0}
\newcommand{\fnresultbestEEs}{0}
\newcommand{\resultaviOOs}{1.95}
\newcommand{\resultaviEEs}{1.51}
\newcommand{\fpresultaviOOs}{0.22}
\newcommand{\fpresultaviEEs}{0.52}
\newcommand{\fnresultaviOOs}{3.34}
\newcommand{\fnresultaviEEs}{0.66}
\newcommand{\resultcvOOs}{1}
\newcommand{\resultcvEEs}{0.28}
\newcommand{\fpresultcvOOs}{31.4}
\newcommand{\fpresultcvEEs}{22.15}
\newcommand{\fnresultcvOOs}{1.6}
\newcommand{\fnresultcvEEs}{1.8}

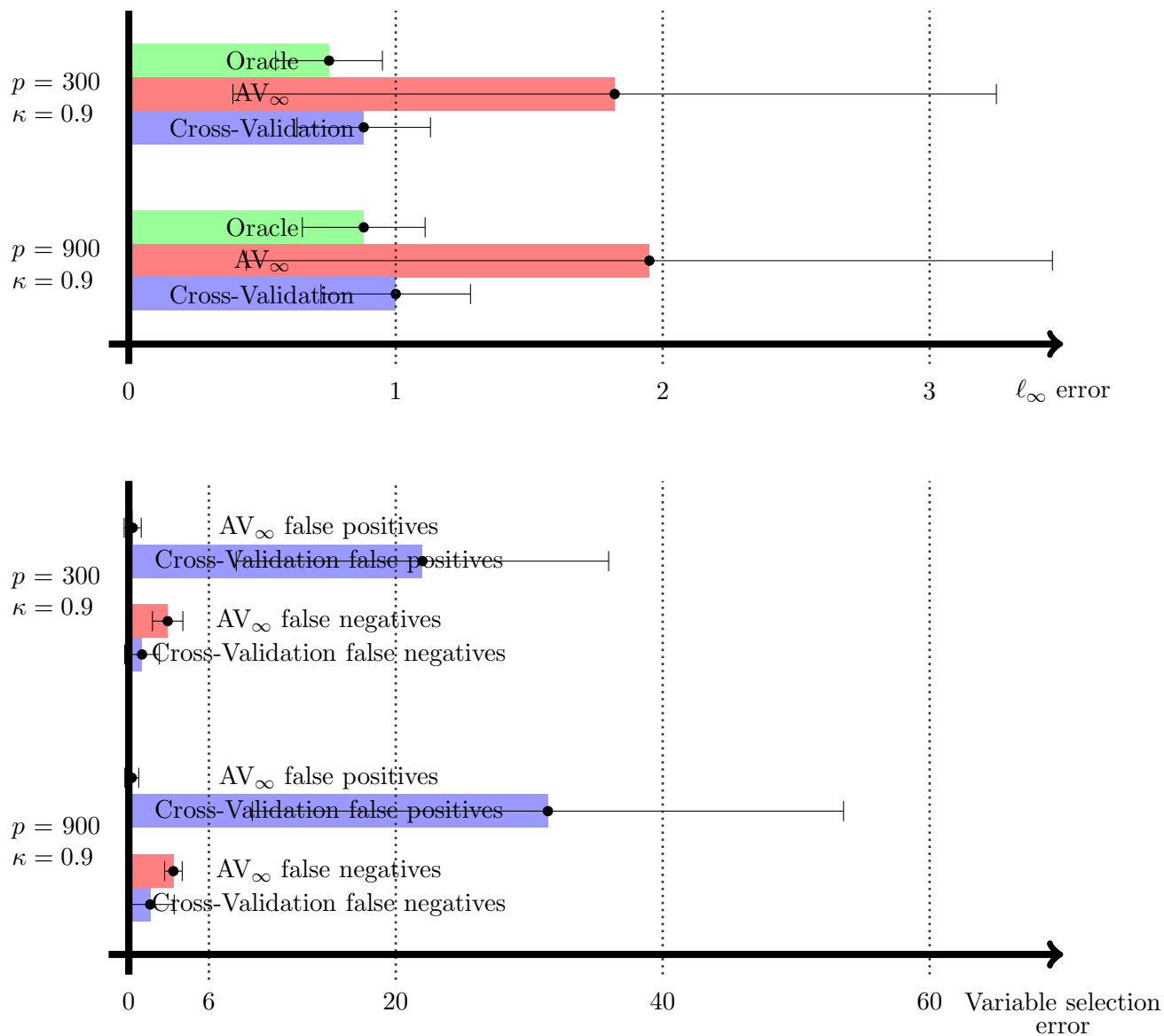
\newcommand{\captionsupstrong}{Sup-norm and variable selection errors
  of the Lasso with three/two different calibration schemes for the
  tuning parameter $\lambda$. Depicted are the results for two
  simulation settings that differ in the number of parameters $p.$ The
  simulation settings and the calibration schemes are specified in the
  main part of the paper.}

\begin{figure}[h]
\begin{tikzpicture}

\path [\colorbest,fill=\colorbest, line width = 0pt] (0,1+\doubleshift) rectangle (\resultbestOOs*\factors,1.5+\doubleshift);
\draw (\resultbestOOs*\factors-\resultbestEEs*\factors,1.25+\doubleshift) -| (\resultbestOOs*\factors+\resultbestEEs*\factors,1.25+\doubleshift);
\draw (\resultbestOOs*\factors-\resultbestEEs*\factors, 1.4+\doubleshift) -- (\resultbestOOs*\factors-\resultbestEEs*\factors, 1.1+\doubleshift);
\draw (\resultbestOOs*\factors+\resultbestEEs*\factors, 1.4+\doubleshift) -- (\resultbestOOs*\factors+\resultbestEEs*\factors, 1.1+\doubleshift);
\filldraw (\resultbestOOs*\factors,1.25+\doubleshift) circle (2pt);

\path [\coloravi,fill=\coloravi, line width = 0pt] (0,0.5+\doubleshift) rectangle (\resultaviOOs*\factors,1+\doubleshift);
\draw (\resultaviOOs*\factors-\resultaviEEs*\factors,0.75+\doubleshift) -| (\resultaviOOs*\factors+\resultaviEEs*\factors,0.75+\doubleshift);
\draw (\resultaviOOs*\factors-\resultaviEEs*\factors, 0.9+\doubleshift) -- (\resultaviOOs*\factors-\resultaviEEs*\factors, 0.6+\doubleshift);
\draw (\resultaviOOs*\factors+\resultaviEEs*\factors, 0.9+\doubleshift) -- (\resultaviOOs*\factors+\resultaviEEs*\factors, 0.6+\doubleshift);
\filldraw (\resultaviOOs*\factors,0.75+\doubleshift) circle (2pt);

\path [\colorcv,fill=\colorcv, line width = 0pt] (0,0+\doubleshift) rectangle (\resultcvOOs*\factors,.5+\doubleshift);
\draw (\resultcvOOs*\factors-\resultcvEEs*\factors,0.25+\doubleshift) -| (\resultcvOOs*\factors+\resultcvEEs*\factors,0.25+\doubleshift);
\draw (\resultcvOOs*\factors-\resultcvEEs*\factors, 0.4+\doubleshift) -- (\resultcvOOs*\factors-\resultcvEEs*\factors, 0.1+\doubleshift);
\draw (\resultcvOOs*\factors+\resultcvEEs*\factors, 0.4+\doubleshift) -- (\resultcvOOs*\factors+\resultcvEEs*\factors, 0.1+\doubleshift);
\filldraw (\resultcvOOs*\factors,0.25+\doubleshift) circle (2pt);

\node[text width = 15mm] at (-1,.7+\doubleshift) {$p=900$ $\correlations=0.9$};

\node at (2,1.25+\doubleshift) {\nameplotoracle};
\node at (2,0.75+\doubleshift) {\avi};
\node at (2,0.25+\doubleshift) {Cross-Validation};

\path [\colorbest,fill=\colorbest, line width = 0pt] (0,1+\shift) rectangle (\resultbestOs*\factors,1.5+\shift);
\draw (\resultbestOs*\factors-\resultbestEs*\factors,1.25+\shift) -| (\resultbestOs*\factors+\resultbestEs*\factors,1.25+\shift);
\draw (\resultbestOs*\factors-\resultbestEs*\factors, 1.4+\shift) -- (\resultbestOs*\factors-\resultbestEs*\factors, 1.1+\shift);
\draw (\resultbestOs*\factors+\resultbestEs*\factors, 1.4+\shift) -- (\resultbestOs*\factors+\resultbestEs*\factors, 1.1+\shift);
\filldraw (\resultbestOs*\factors,1.25+\shift) circle (2pt);

\path [\coloravi,fill=\coloravi, line width = 0pt] (0,0.5+\shift) rectangle (\resultaviOs*\factors,1+\shift);
\draw (\resultaviOs*\factors-\resultaviEs*\factors,0.75+\shift) -| (\resultaviOs*\factors+\resultaviEs*\factors,0.75+\shift);
\draw (\resultaviOs*\factors-\resultaviEs*\factors, 0.9+\shift) -- (\resultaviOs*\factors-\resultaviEs*\factors, 0.6+\shift);
\draw (\resultaviOs*\factors+\resultaviEs*\factors, 0.9+\shift) -- (\resultaviOs*\factors+\resultaviEs*\factors, 0.6+\shift);
\filldraw (\resultaviOs*\factors,0.75+\shift) circle (2pt);

\path [\colorcv,fill=\colorcv, line width = 0pt] (0,0+\shift) rectangle (\resultcvOs*\factors,.5+\shift);
\draw (\resultcvOs*\factors-\resultcvEs*\factors,0.25+\shift) -| (\resultcvOs*\factors+\resultcvEs*\factors,0.25+\shift);
\draw (\resultcvOs*\factors-\resultcvEs*\factors, 0.4+\shift) -- (\resultcvOs*\factors-\resultcvEs*\factors, 0.1+\shift);
\draw (\resultcvOs*\factors+\resultcvEs*\factors, 0.4+\shift) -- (\resultcvOs*\factors+\resultcvEs*\factors, 0.1+\shift);
\filldraw (\resultcvOs*\factors,0.25+\shift) circle (2pt);

\node[text width = 15mm] at (-1,.7+\shift) {$p=300$ $\correlations=0.9$};

\node at (2,1.25+\shift) {\nameplotoracle};
\node at (2,0.75+\shift) {\avi};
\node at (2,0.25+\shift) {Cross-Validation};

\draw[line width = 1pt, black!80, dotted] (1*\factors,-5.8) --(1*\factors,-0.5);
\draw[line width = 1pt, black!80, dotted] (2*\factors,-5.8) --(2*\factors,-0.5);
\draw[line width = 1pt, black!80, dotted] (3*\factors,-5.8) --(3*\factors,-0.5);

\draw[line width = 3pt] (-0,-5.8) --(0,-0.5); 
\draw[line width = 3pt, ->] (-0.3,-5.5) --(14,-5.5); 
\node at (14,-6.2) {$\ell_\infty$ error}; 
\node at (0,-6.2) {$0$};
\node at (1*\factors,-6.2) {$1$};
\node at (2*\factors,-6.2) {$2$};
\node at (3*\factors,-6.2) {$3$};

\end{tikzpicture}

~\vspace{5mm}

\begin{tikzpicture}

\path [\coloravi,fill=\coloravi, line width = 0pt] (0,1.9+\vsdoubleshift) rectangle (\fpresultaviOOs*\vsfactor,2.4+\vsdoubleshift);
\draw (\fpresultaviOOs*\vsfactor-\fpresultaviEEs*\vsfactor,2.15+\vsdoubleshift) -| (\fpresultaviOOs*\vsfactor+\fpresultaviEEs*\vsfactor,2.15+\vsdoubleshift);
\draw (\fpresultaviOOs*\vsfactor-\fpresultaviEEs*\vsfactor, 2.3+\vsdoubleshift) -- (\fpresultaviOOs*\vsfactor-\fpresultaviEEs*\vsfactor, 2+\vsdoubleshift);
\draw (\fpresultaviOOs*\vsfactor+\fpresultaviEEs*\vsfactor, 2.3+\vsdoubleshift) -- (\fpresultaviOOs*\vsfactor+\fpresultaviEEs*\vsfactor, 2+\vsdoubleshift);
\filldraw (\fpresultaviOOs*\vsfactor,2.15+\vsdoubleshift) circle (2pt);
\path [\colorcv,fill=\colorcv, line width = 0pt] (0,1.4+\vsdoubleshift) rectangle (\fpresultcvOOs*\vsfactor,1.9+\vsdoubleshift);
\draw (\fpresultcvOOs*\vsfactor-\fpresultcvEEs*\vsfactor,1.65+\vsdoubleshift) -| (\fpresultcvOOs*\vsfactor+\fpresultcvEEs*\vsfactor,1.65+\vsdoubleshift);
\draw (\fpresultcvOOs*\vsfactor-\fpresultcvEEs*\vsfactor, 1.8+\vsdoubleshift) -- (\fpresultcvOOs*\vsfactor-\fpresultcvEEs*\vsfactor, 1.5+\vsdoubleshift);
\draw (\fpresultcvOOs*\vsfactor+\fpresultcvEEs*\vsfactor, 1.8+\vsdoubleshift) -- (\fpresultcvOOs*\vsfactor+\fpresultcvEEs*\vsfactor, 1.5+\vsdoubleshift);
\filldraw (\fpresultcvOOs*\vsfactor,1.65+\vsdoubleshift) circle (2pt);

\path [\coloravi,fill=\coloravi, line width = 0pt] (0,0.5+\vsdoubleshift) rectangle (\fnresultaviOOs*\vsfactor,1+\vsdoubleshift);
\draw (\fnresultaviOOs*\vsfactor-\fnresultaviEEs*\vsfactor,0.75+\vsdoubleshift) -| (\fnresultaviOOs*\vsfactor+\fnresultaviEEs*\vsfactor,0.75+\vsdoubleshift);
\draw (\fnresultaviOOs*\vsfactor-\fnresultaviEEs*\vsfactor, 0.9+\vsdoubleshift) -- (\fnresultaviOOs*\vsfactor-\fnresultaviEEs*\vsfactor, 0.6+\vsdoubleshift);
\draw (\fnresultaviOOs*\vsfactor+\fnresultaviEEs*\vsfactor, 0.9+\vsdoubleshift) -- (\fnresultaviOOs*\vsfactor+\fnresultaviEEs*\vsfactor, 0.60+\vsdoubleshift);
\filldraw (\fnresultaviOOs*\vsfactor,0.75+\vsdoubleshift) circle (2pt);
\path [\colorcv,fill=\colorcv, line width = 0pt] (0,-0.0+\vsdoubleshift) rectangle (\fnresultcvOOs*\vsfactor,0.5+\vsdoubleshift);
\draw (\fnresultcvOOs*\vsfactor-\fnresultcvEEs*\vsfactor,0.25+\vsdoubleshift) -| (\fnresultcvOOs*\vsfactor+\fnresultcvEEs*\vsfactor,0.25+\vsdoubleshift);
\draw (\fnresultcvOOs*\vsfactor-\fnresultcvEEs*\vsfactor, 0.4+\vsdoubleshift) -- (\fnresultcvOOs*\vsfactor-\fnresultcvEEs*\vsfactor, 0.1+\vsdoubleshift);
\draw (\fnresultcvOOs*\vsfactor+\fnresultcvEEs*\vsfactor, 0.4+\vsdoubleshift) -- (\fnresultcvOOs*\vsfactor+\fnresultcvEEs*\vsfactor, 0.1+\vsdoubleshift);
\filldraw (\fnresultcvOOs*\vsfactor,0.25+\vsdoubleshift) circle (2pt);

\node[text width = 15mm] at (-1,1.2+\vsdoubleshift) {$p=900$ $\correlations=0.9$};

\node at (3,2.15+\vsdoubleshift) {\avi\ false positives};
\node at (3,1.65+\vsdoubleshift) {Cross-Validation false positives};
\node at (3,0.75+\vsdoubleshift) {\avi\ false negatives};
\node at (3,0.25+\vsdoubleshift) {Cross-Validation false negatives};

\path [\coloravi,fill=\coloravi, line width = 0pt] (0,1.9+\vsshift) rectangle (\fpresultaviOs*\vsfactor,2.4+\vsshift);
\draw (\fpresultaviOs*\vsfactor-\fpresultaviEs*\vsfactor,2.15+\vsshift) -| (\fpresultaviOs*\vsfactor+\fpresultaviEs*\vsfactor,2.15+\vsshift);
\draw (\fpresultaviOs*\vsfactor-\fpresultaviEs*\vsfactor, 2.3+\vsshift) -- (\fpresultaviOs*\vsfactor-\fpresultaviEs*\vsfactor, 2+\vsshift);
\draw (\fpresultaviOs*\vsfactor+\fpresultaviEs*\vsfactor, 2.3+\vsshift) -- (\fpresultaviOs*\vsfactor+\fpresultaviEs*\vsfactor, 2+\vsshift);
\filldraw (\fpresultaviOs*\vsfactor,2.15+\vsshift) circle (2pt);
\path [\colorcv,fill=\colorcv, line width = 0pt] (0,1.4+\vsshift) rectangle (\fpresultcvOs*\vsfactor,1.9+\vsshift);
\draw (\fpresultcvOs*\vsfactor-\fpresultcvEs*\vsfactor,1.65+\vsshift) -| (\fpresultcvOs*\vsfactor+\fpresultcvEs*\vsfactor,1.65+\vsshift);
\draw (\fpresultcvOs*\vsfactor-\fpresultcvEs*\vsfactor, 1.8+\vsshift) -- (\fpresultcvOs*\vsfactor-\fpresultcvEs*\vsfactor, 1.5+\vsshift);
\draw (\fpresultcvOs*\vsfactor+\fpresultcvEs*\vsfactor, 1.8+\vsshift) -- (\fpresultcvOs*\vsfactor+\fpresultcvEs*\vsfactor, 1.5+\vsshift);
\filldraw (\fpresultcvOs*\vsfactor,1.65+\vsshift) circle (2pt);

\path [\coloravi,fill=\coloravi, line width = 0pt] (0,0.5+\vsshift) rectangle (\fnresultaviOs*\vsfactor,1+\vsshift);
\draw (\fnresultaviOs*\vsfactor-\fnresultaviEs*\vsfactor,0.75+\vsshift) -| (\fnresultaviOs*\vsfactor+\fnresultaviEs*\vsfactor,0.75+\vsshift);
\draw (\fnresultaviOs*\vsfactor-\fnresultaviEs*\vsfactor, 0.9+\vsshift) -- (\fnresultaviOs*\vsfactor-\fnresultaviEs*\vsfactor, 0.6+\vsshift);
\draw (\fnresultaviOs*\vsfactor+\fnresultaviEs*\vsfactor, 0.9+\vsshift) -- (\fnresultaviOs*\vsfactor+\fnresultaviEs*\vsfactor, 0.60+\vsshift);
\filldraw (\fnresultaviOs*\vsfactor,0.75+\vsshift) circle (2pt);
\path [\colorcv,fill=\colorcv, line width = 0pt] (0,-0.0+\vsshift) rectangle (\fnresultcvOs*\vsfactor,0.5+\vsshift);
\draw (\fnresultcvOs*\vsfactor-\fnresultcvEs*\vsfactor,0.25+\vsshift) -| (\fnresultcvOs*\vsfactor+\fnresultcvEs*\vsfactor,0.25+\vsshift);
\draw (\fnresultcvOs*\vsfactor-\fnresultcvEs*\vsfactor, 0.4+\vsshift) -- (\fnresultcvOs*\vsfactor-\fnresultcvEs*\vsfactor, 0.1+\vsshift);
\draw (\fnresultcvOs*\vsfactor+\fnresultcvEs*\vsfactor, 0.4+\vsshift) -- (\fnresultcvOs*\vsfactor+\fnresultcvEs*\vsfactor, 0.1+\vsshift);
\filldraw (\fnresultcvOs*\vsfactor,0.25+\vsshift) circle (2pt);

\node[text width = 15mm] at (-1,1.2+\vsshift) {$p=300$ $\correlations=0.9$};

\node at (3,2.15+\vsshift) {\avi\ false positives};
\node at (3,1.65+\vsshift) {Cross-Validation false positives};
\node at (3,0.75+\vsshift) {\avi\ false negatives};
\node at (3,0.25+\vsshift) {Cross-Validation false negatives};

\draw[line width = 1pt, black!80, dotted] (20*\vsfactor,-5.88) --(20*\vsfactor,1.6);
\draw[line width = 1pt, black!80, dotted] (40*\vsfactor,-5.8) --(40*\vsfactor,1.6);
\draw[line width = 1pt, black!80, dotted] (60*\vsfactor,-5.8) --(60*\vsfactor,1.6);
\draw[line width = 3pt] (-0,-5.8) --(0,1.6); 
\draw[line width = 3pt, ->] (-0.3,-5.5) --(14,-5.5); 
\node at (14,-6.2) {Variable selection}; 
\node at (14,-6.6) {error}; 
\node at (0,-6.2) {$0$};
\node at (20*\vsfactor,-6.2) {$20$};
\node at (40*\vsfactor,-6.2) {$40$};
\node at (60*\vsfactor,-6.2) {$60$};
 
\node at (6*\vsfactor,-6.2) {$6$};
\draw[line width = 1pt, black!80, dotted] (6*\vsfactor,-5.88) --(6*\vsfactor,1.6);

\draw[line width = 3pt] (-0,-5.8) --(0,1.6); 
\draw[line width = 3pt, ->] (-0.3,-5.5) --(14,-5.5); 

\end{tikzpicture}
\caption{\captionsupstrong}
\label{MainComparisonStrong}
\end{figure}


\newcommand{\captionVSstrong}{Number of false positives $|\{j:\bt_j= 0,(\widehat\beta_\lambda)_j\neq 0\}|$ and false negatives $|\{j:\bt_j\neq 0,(\widehat\beta_\lambda)_j= 0\}|$ of the Lasso with \avi\ and Cross-Validation as calibration schemes for the tuning parameter~$\lambda$. For \avi, the safe threshold described after Theorem~\ref{ThmAvi} is applied. The simulations settings correspond to those in Figure~\ref{MainComparisonStrong}.}

\revision{\section{Strong Correlations} In this paper, we assume that
  the correlations in design matrix are small, which is needed for
  precise $\ell_\infty$-estimation and variable selection. In the
  interest of completeness, however, we add here two simulations where
  the correlations are large. Overall, we use the same settings as
  described in the main part of the paper, but we
  set~$\kappa=0.9$. The results are summarized in
  Figure~\ref{MainComparisonStrong} (note that the x-scale in the
  upper part of the figure is different from the scales of the
  corresponding plots in the main part of the paper). We find that
  \avi\ misses about half of the pertinent variables but has almost no
  false positives. Cross-Validation, on the other hand, has less false
  negatives but selects many irrelevant variables. As expected, none
  of the methods, including the oracle, provide accurate
  $\ell_\infty$-estimation.}



\vskip 0.2in
\bibliography{Literature}

\begin{thebibliography}{36}
\providecommand{\natexlab}[1]{#1}
\providecommand{\url}[1]{\texttt{#1}}
\expandafter\ifx\csname urlstyle\endcsname\relax
  \providecommand{\doi}[1]{doi: #1}\else
  \providecommand{\doi}{doi: \begingroup \urlstyle{rm}\Url}\fi

\bibitem[Antoniadis(2010)]{Antoniadis10}
A.~Antoniadis.
\newblock Comments on: {$\ell_1$}-penalization for mixture regression models.
\newblock \emph{Test}, 19\penalty0 (2):\penalty0 257--258, 2010.

\bibitem[Bach(2008)]{Bach08}
F.~Bach.
\newblock Bolasso: model consistent lasso estimation through the bootstrap.
\newblock In \emph{Proceedings of the 25th International Conference on Machine
  Learning}, pages 33--40, 2008.

\bibitem[Baraud et~al.(2014)Baraud, Giraud, and Huet]{Baraud14}
Y.~Baraud, C.~Giraud, and S.~Huet.
\newblock Estimator selection in the gaussian setting.
\newblock In \emph{Ann. Inst. H. Poincar\'e Probab. Statist.}, volume~50, pages
  1092--1119, 2014.

\bibitem[Belloni et~al.(2011)Belloni, Chernozhukov, and Wang]{Belloni11}
A.~Belloni, V.~Chernozhukov, and L.~Wang.
\newblock Square-root lasso: pivotal recovery of sparse signals via conic
  programming.
\newblock \emph{Biometrika}, 98\penalty0 (4):\penalty0 791--806, 2011.

\bibitem[Bickel et~al.(2009)Bickel, Ritov, and Tsybakov]{Bickel09}
P.~Bickel, Y.~Ritov, and A.~Tsybakov.
\newblock Simultaneous analysis of the {L}asso and {D}antzig selector.
\newblock \emph{Ann. Statist.}, 37\penalty0 (4):\penalty0 1705--1732, 2009.

\bibitem[B{\"u}hlmann and van~de Geer(2011)]{Buhlmann11}
P.~B{\"u}hlmann and S.~van~de Geer.
\newblock \emph{Statistics for high-dimensional data: Methods, theory and
  applications}.
\newblock Springer Series in Statistics. Springer, 2011.

\bibitem[B\"uhlmann et~al.(2014)B\"uhlmann, Kalisch, and Meier]{Buhlmann:2014}
P.~B\"uhlmann, M.~Kalisch, and L.~Meier.
\newblock High-dimensional statistics with a view toward applications in
  biology.
\newblock \emph{Annual Review of Statistics and Its Application}, 1\penalty0
  (1):\penalty0 255--278, 2014.

\bibitem[Bunea(2008)]{BuneaEN}
F.~Bunea.
\newblock Honest variable selection in linear and logistic regression models
  via {$\ell_1$} and {$\ell_1+\ell_2$} penalization.
\newblock \emph{Electron. J. Stat.}, 2:\penalty0 1153--1194, 2008.

\bibitem[Ch{\'e}telat et~al.(2014)Ch{\'e}telat, Lederer, and
  Salmon]{Chetelat14}
D.~Ch{\'e}telat, J.~Lederer, and J.~Salmon.
\newblock Optimal two-step prediction in regression.
\newblock \emph{arXiv:1410.5014}, 2014.

\bibitem[Chichignoud and Lederer(2014)]{ChichiYoyo12}
M.~Chichignoud and J.~Lederer.
\newblock A robust, adaptive {M}-estimator for pointwise estimation in
  heteroscedastic regression.
\newblock \emph{Bernoulli}, 20\penalty0 (3):\penalty0 1560--1599, 2014.

\bibitem[Dalalyan et~al.(2014)Dalalyan, Hebiri, and Lederer]{ArnakMoYo14}
A.~Dalalyan, M.~Hebiri, and J.~Lederer.
\newblock On the {P}rediction {P}erformance of the {L}asso.
\newblock \emph{Bernoulli, in press}, 2014.

\bibitem[Fan and Li(2001)]{Fan_Li01}
J.~Fan and R.~Li.
\newblock Variable selection via nonconcave penalized likelihood and its oracle
  properties.
\newblock \emph{J. Amer. Statist. Assoc.}, 96\penalty0 (456):\penalty0
  1348--1360, 2001.

\bibitem[Friedman et~al.(2010)Friedman, Hastie, and Tibshirani]{glmnet10}
J.~Friedman, T.~Hastie, and R.~Tibshirani.
\newblock Regularization paths for generalized linear models via coordinate
  descent.
\newblock \emph{Journal of Statistical Software}, 33\penalty0 (1):\penalty0
  1--22, 2010.

\bibitem[Giraud et~al.(2012)Giraud, Huet, and Verzelen]{Giraud12}
C.~Giraud, S.~Huet, and N.~Verzelen.
\newblock High-dimensional regression with unknown variance.
\newblock \emph{Statistical Science}, 27\penalty0 (4):\penalty0 500--518, 2012.

\bibitem[Hebiri and Lederer(2013)]{YoyoMomo12}
M.~Hebiri and J.~Lederer.
\newblock How correlations influence {L}asso prediction.
\newblock \emph{IEEE Trans. Inform. Theory}, 59\penalty0 (3):\penalty0
  1846--1854, 2013.

\bibitem[Lederer and M{\"u}ller(2015)]{Lederer:14}
J.~Lederer and C.~M{\"u}ller.
\newblock Don't fall for tuning parameters: Tuning-free variable selection in
  high dimensions with the trex.
\newblock In \emph{Proceedings of the Twenty-Ninth AAAI Conference on
  Artificial Intelligence}, 2015.

\bibitem[Leeb and P{\"o}tscher(2008)]{Leeb08}
H.~Leeb and B.~P{\"o}tscher.
\newblock Sparse estimators and the oracle property, or the return of {H}odges'
  estimator.
\newblock \emph{J. Econometrics}, 142\penalty0 (1):\penalty0 201--211, 2008.

\bibitem[Lepski(1990)]{Lepski90}
O.~Lepski.
\newblock A problem of adaptive estimation in {G}aussian white noise.
\newblock \emph{Teor. Veroyatnost. i Primenen.}, 35\penalty0 (3):\penalty0
  459--470, 1990.
\newblock ISSN 0040-361X.

\bibitem[Lepski et~al.(1997)Lepski, Mammen, and
  Spokoiny]{Lepski_Mammen_Spokoiny97}
O.~Lepski, E.~Mammen, and V.~Spokoiny.
\newblock Optimal spatial adaptation to inhomogeneous smoothness: an approach
  based on kernel estimates with variable bandwidth selectors.
\newblock \emph{Ann. Statist.}, 25\penalty0 (3):\penalty0 929--947, 1997.

\bibitem[Loh and Wainwright(2014)]{Loh14}
P.-L. Loh and M.~Wainwright.
\newblock Support recovery without incoherence: A case for nonconvex
  regularization.
\newblock \emph{arXiv:1412.5632}, 2014.

\bibitem[Lounici(2008)]{Lounici08}
K.~Lounici.
\newblock Sup-norm convergence rate and sign concentration property of {L}asso
  and {D}antzig estimators.
\newblock \emph{Electron. J. Stat.}, 2:\penalty0 90--102, 2008.

\bibitem[Meinshausen and B{\"u}hlmann(2010)]{Meinshausen10}
N.~Meinshausen and P.~B{\"u}hlmann.
\newblock Stability selection.
\newblock \emph{J. R. Stat. Soc. Ser. B. Stat. Methodol.}, 72\penalty0
  (4):\penalty0 417--473, 2010.

\bibitem[Negahban et~al.(2012)Negahban, Ravikumar, Wainwright, and
  Yu]{NegRavWaiYu12}
S.~Negahban, P.~Ravikumar, M.~J. Wainwright, and B.~Yu.
\newblock A unified framework for high-dimensional analysis of {$M$}-estimators
  with decomposable regularizers.
\newblock \emph{Statistical Science}, 27\penalty0 (4):\penalty0 538--557,
  December 2012.

\bibitem[{R Core Team}(2013)]{Rsoftware}
{R Core Team}.
\newblock \emph{R: A Language and Environment for Statistical Computing}.
\newblock R Foundation for Statistical Computing, Vienna, Austria, 2013.
\newblock http://www.R-project.org/.

\bibitem[Sabourin et~al.(2015)Sabourin, Valdar, and Nobel]{Sabourin15}
J.~Sabourin, W.~Valdar, and A.~Nobel.
\newblock A permutation approach for selecting the penalty parameter in
  penalized model selection.
\newblock \emph{Biometrics}, 71\penalty0 (4):\penalty0 1185--1194, 2015.

\bibitem[St{\"a}dler et~al.(2010)St{\"a}dler, B{\"u}hlmann, and {van de
  Geer}]{Stadler10}
N.~St{\"a}dler, P.~B{\"u}hlmann, and S.~{van de Geer}.
\newblock $\ell_{1}$-penalization for mixture regression models.
\newblock \emph{Test}, 19\penalty0 (2):\penalty0 209--256, 2010.

\bibitem[Sun and Zhang(2012)]{ScaledLasso11}
T.~Sun and {C.-H.} Zhang.
\newblock Scaled sparse linear regression.
\newblock \emph{Biometrika}, 99\penalty0 (4):\penalty0 879--898, 2012.

\bibitem[Tibshirani(1996)]{Tibshirani-LASSO}
R.~Tibshirani.
\newblock Regression shrinkage and selection via the lasso.
\newblock \emph{J. Roy. Statist. Soc. Ser. B}, 58\penalty0 (1):\penalty0
  267--288, 1996.

\bibitem[van~de Geer(2007)]{vandeG07}
S.~van~de Geer.
\newblock The deterministic {L}asso.
\newblock \emph{2007 Proc. Amer. Math. Soc.
  [CD-ROM],~see~also~www.stat.math.ethz.ch/\textasciitilde geer/lasso.pdf},
  2007.

\bibitem[van~de Geer(2014)]{vdGeer14}
S.~van~de Geer.
\newblock Worst possible sub-directions in high-dimensional models.
\newblock \emph{J. Multivariate Anal., in press}, 2014.

\bibitem[van~de Geer and B{\"u}hlmann(2009)]{Sara09}
S.~van~de Geer and P.~B{\"u}hlmann.
\newblock On the conditions used to prove oracle results for the {L}asso.
\newblock \emph{Electron. J. Stat.}, 3:\penalty0 1360--1392, 2009.

\bibitem[van~de Geer and Lederer(2013)]{vdGeer11}
S.~van~de Geer and J.~Lederer.
\newblock The {L}asso, correlated design, and improved oracle inequalities.
\newblock \emph{IMS Collections}, 9:\penalty0 303--316, 2013.

\bibitem[Yang(2005)]{Yang05}
Y.~Yang.
\newblock Can the strengths of aic and bic be shared? {A} conflict between
  model indentification and regression estimation.
\newblock \emph{Biometrika}, 92\penalty0 (4):\penalty0 937--950, 2005.

\bibitem[Ye and Zhang(2010)]{YeZhang10}
F.~Ye and C.-H. Zhang.
\newblock Rate minimaxity of the lasso and dantzig selector for the lq loss in
  lr balls.
\newblock \emph{J. Mach. Learn. Res.}, 11:\penalty0 3519--3540, 2010.

\bibitem[Zhang(2010)]{Zhang10}
C.-H. Zhang.
\newblock Nearly unbiased variable selection under minimax concave penalty.
\newblock \emph{Ann. Statist.}, pages 894--942, 2010.

\bibitem[Zhao and Yu(2006)]{Zhao06}
P.~Zhao and B.~Yu.
\newblock On model selection consistency of {L}asso.
\newblock \emph{J. Mach. Learn. Res.}, 7:\penalty0 2541--2563, 2006.

\end{thebibliography}

\end{document}